\documentclass[aps,prc,preprint,tightenlines,groupedaddress,floatfix,showpacs]{revtex4-1}
\usepackage{amsmath,amssymb,latexsym,array,bm}
\usepackage{graphicx}

\begin{document}

\title{\emph{R}-Matrix description of particle energy spectra produced by
  low-energy $\text{T}+\text{T}$ reactions}

\author{C.~R.~Brune}
\affiliation{%
  Edwards Accelerator Laboratory \\
  Department of Physics and Astronomy \\
  Ohio University, Athens, Ohio 45701, USA }
\author{J.~A.~Caggiano}
\author{D.~B.~Sayre}
\affiliation{%
  Lawrence Livermore National Laboratory \\
  Livermore, California 94550, USA }
\author{A.~D.~Bacher}
\affiliation{%
  Indiana University Cyclotron Facility (IUCF) \\
   Bloomington, Indiana 47408, USA }
\author{G.~M.~Hale}
\author{M.~W.~Paris}
\affiliation{%
  Los Alamos National Laboratory \\
  Los Alamos, New Mexico 87545, USA }
\date{\today}

\begin{abstract}
An $R$-matrix model for three-body final states is presented and
applied to a recent measurement of the neutron energy spectrum
from the $\text{T}+\text{T}\rightarrow 2n+\alpha$ reaction.
The calculation includes the $n\alpha$ and $nn$ interactions in the
final state, angular momentum conservation, antisymmetrization, and the
interference between different channels.
A good fit to the measured spectrum is obtained, where clear evidence
for the ${}^5{\rm He}$ ground state is observed.
The model is also used to predict the $\alpha$-particle spectrum
from $\text{T}+\text{T}$ as well as particle spectra from
${}^3\text{He}+{}^3\text{He}$.
The $R$-matrix approach presented here is very general, and can be
adapted to a wide variety of problems with three-body final states.
\end{abstract}


\maketitle

\section{Introduction}
\label{sec:intro}

Due to the presence of three particles in the final state, the
$\text{T}+\text{T}\rightarrow 2n+\alpha$ reaction produces distributions
of neutron and $\alpha$-particle energies.
The neutron energy spectrum at an effective $E_{c.m.}$ of 16~keV has
recently been measured in an inertial confinement fusion experiment
at the National Ignition Facility (NIF)~\cite{Say13}.
This paper also presented a sequential-decay $R$-matrix model
for the three-body state.
The primary purpose of the present paper is to fully describe this model and
to explore a broader range of assumptions for the fitting of the
neutron spectrum. We also present a prediction for the $\alpha$-particle
spectrum, for which limited data exists. Finally, we calculate
the final-state energy spectra of the mirror reaction
${}^3\text{He}+{}^3\text{He}$ and discuss some features of our approach when
one of the nuclei in the final state is much heavier than the others.

Our model includes interactions between all pairs of nuclei
in the final state. For the $\text{T}+\text{T}$ case, this implies the
$n\alpha$ interaction, including the unbound $3/2^-$ ground and $1/2^-$ first
excited state of ${}^5{\rm He}$, and the $nn$ interaction.
The calculation also incorporates angular momentum conservation and
fermion symmetry. The latter is a particular example of an order-of-emission
effect, which give rise to various interference phenomena.
In addition, kinematic effects present in the
three-body final state are tightly integrated into the model.
These details of the model, predictions for particle spectra, and
comparisons to available experimental data are discussed below.

The reactions ${}^3\text{He}+{}^3\text{He}\rightarrow 2p+\alpha$ and
$\text{T}+{}^3\text{He}\rightarrow n+p+\alpha$, which are
related by mirror or isospin symmetry to
$\text{T}+\text{T}\rightarrow 2n+\alpha$, are also presently under
study at inertial confinement fusion facilities.
The model presented here can be adapted to these reactions, and a
prediction for the ${}^3\text{He}+{}^3\text{He}\rightarrow 2p+\alpha$
case is given in this paper.
This $R$-matrix approach is very general, and additional areas where
it could be applied are discussed in the conclusion.

\section{Three-body kinematics}
\label{sec:kinematics}

The three-body final state from the $\text{T}+\text{T}\rightarrow 2n+\alpha$
reaction will be described using non-relativistic kinematics~\cite{Ohl65}.
With the center of mass assumed to be at rest, the kinetic energy
available in the final state is given by
\begin{equation}
E_{tot} = Q+E_{c.m.},
\end{equation}
where $Q=11.332$~MeV and $E_{c.m.}$ is the center-of-mass (c.m.)
kinetic energy of the initial state.
Here, we will assume $E_{c.m.}=16$~keV, unless otherwise indicated.
The masses of the final-state particles
are $m_i$, where $i=1$, 2, or 3, and $M=m_1+m_2+m_3$.
Indices 1~and~2 are used for the neutrons, with index~3 used for
the $\alpha$ particle.
The momentum and kinetic energies of the final-state particles are given
in the three-body c.m. system by
$\bm{p}_i$ and $E_i$.
The relative momentum and kinetic energy of particles $i$~and~$j$ are
\begin{eqnarray}
\bm{p}_{ij} &=& \frac{\mu_{ij}}{m_i} \bm{p}_i
  - \frac{\mu_{ij}}{m_j} \bm{p}_j \quad {\rm and}
  \label{eq:momentum_1} \\
E_{ij} &=& \frac{p_{ij}^2}{2\mu_{ij}}, \quad {\rm where} \\
\mu_{ij} &=& \frac{m_i m_j}{m_i+m_j}.
\end{eqnarray}
Assuming the indices $i$, $j$, and $k$ are all distinct,
the relative kinetic energy between particle~$i$
and the $j-k$ system is given by $\tilde{E_i}$ and we also have
\begin{eqnarray}
0 &=& \bm{p}_1+\bm{p}_2+\bm{p}_3,
  \label{eq:momentum_2} \\
\tilde{E_i} &=& \frac{M}{m_j+m_k} E_i, \quad {\rm and} \\
E_{tot} &=& E_{1}+E_{2}+E_{3}=\tilde{E_i}+E_{jk}. \label{eq:etot}
\end{eqnarray}

\section{\emph{R}-matrix model}
\label{sec:rmatrix}

The energy distribution of particles emitted by reactions proceeding to
unbound states can be described using $R$-matrix methods, as presented by
Baker~\cite{Bar88}.
This approach in essence describes the particle emissions as sequential
two-body decays.
Due to the low energy in the initial $\text{T}+\text{T}$ state, 
we assume it has orbital angular momentum of zero and thus a
total spin and parity of $0^+$.
We consider here two types of sequential decays: neutron emission to
unbound ${}^5{\rm He}$ intermediate states and $\alpha$-particle emission
to unbound neutron--neutron states. This latter type decay may
also be referred to as di-neutron emission~\cite{Lac65}.
For both of these decay types, a further complication is presented by
the fact that the amplitudes must be constructed to be antisymmetric under
the exchange of neutrons, which give rise to {\em direct} and
{\em exchange} terms [see Eq.~\eqref{eqn:dir} and Eq.~\eqref{eqn:exch}, 
respectively].
The $R$-matrix formalism has been applied to other cases of three-body final
states with identical particles in Refs.~\cite{Bal74,Gee77,Fyn03,Fyn09}.

We emphasize that this approach treats the intermediate
state rather carefully. The phase shifts between the particles that make up
this state are rather well known for the cases considered here -- either
experimentally ($n\alpha$ scattering) or theoretically ($nn$ scattering).
The $R$-matrix model described below accurately incorporates these
phase shifts.
The interaction between the first particle emitted and
the intermediate state is, however, not well known, as it generally
cannot be studied independently in the laboratory.
This part of the matrix element is treated in a minimalist $R$-matrix
approach, with just a hard-sphere interaction, which characterizes
a non-resonant phase shift.

\subsection{Neutron emission through ${}^5{\rm He}$ intermediate states}
\label{subsec:he5}

For our assumption of a $0^+$ initial state, both neutrons must have
the identical orbital angular momentum $l$.
We assume the amplitude for the process is given by
\begin{equation}
\mathcal{M}_{\nu_1\nu_2} = \sum_c u_c(\tilde{E_1})
  f^{lJ}_{\nu_1\nu_2}(\Omega_1,\Omega_{23}),
\end{equation}
where $\nu_i$ are the spin projections of the neutrons
and the energy dependence is described by the $R$-matrix expression
\begin{equation}
u_c(\tilde{E_1})=\left[\frac{P_1 P_{23}}{p_1 p_{23}}\right]^{1/2}
  e^{i(\omega_1-\Phi_1)} e^{i(\omega_{23}-\Phi_{23})}
  \frac{\sum_\lambda\frac{A_{c\lambda}\gamma_{c\lambda}}{E_{c\lambda}-E_{23}}}
  {1-[S_{23}-B_c+iP_{23}]R_c}
\label{eq:u_n_alpha}
\end{equation}
and the spin and angle dependence is given described by
$f^{lJ}_{\nu_1\nu_2}(\Omega_1,\Omega_{23})$.
The subscripts 1, 2, and~3 refer to the first neutron emitted, second neutron
emitted, and the $\alpha$~particle, respectively.
The channel is labeled by $c\equiv(l,J,\beta)$, where $J$ is the
angular momentum of the intermediate state and
$\beta$ indicates the decay type which is via
${}^5{\rm He}$ intermediate states in this case ($\beta=n\alpha$).
The quantity $R_c$ is the $n+\alpha$ elastic-scattering $R$~matrix
\begin{equation}
R_c=\sum_\lambda \frac{\gamma_{c\lambda}^2}{E_{c\lambda}-E_{23}}
\end{equation}
and $E_{c\lambda}$, $\gamma_{c\lambda}$, $A_{c\lambda}$, and $B_c$ are
the $R$-matrix parameters: the level energies, reduced width amplitudes,
feeding factors, and boundary-condition constants, respectively.
The $R$-matrix surface functions depend upon the channel radii,
$l$, and energy and include the penetration factors $P_1$ and $P_{23}$,
the shift function $S_{23}$,
and the hard-sphere phase shifts $-\Phi_1$ and $-\Phi_{23}$.
The quantities $\omega_1$ and $\omega_{23}$  are the Coulomb phase shifts
which are zero in this case.
Note that the penetration factors have been divided by the
corresponding momentum.
This convention is also used in Refs.~\cite{Bal74,Gee77} and removes two-body
phase space factors present in the penetration factors from the
three-body matrix element~\cite{Fyn03}.

The spin and angle dependences are calculated by first coupling a
neutron with spin projection $\nu_2$ to an $\alpha$ particle to form a
${}^5{\rm He}$ state with angular momentum quantum numbers $(J,m_J)$
\begin{equation}
g^{lJ}_{\nu_2,m_J}(\Omega_{23}) = \sum_{m,m_l}
  \langle lm_l\frac{1}{2}\nu_2|J m_J-m\rangle Y_{lm_l}(\bm{\hat{p}}_{23})
\end{equation}
and then coupling to another neutron with spin projection $\nu_1$
to form the $0^+$ $\text{T}+\text{T}$ state
\begin{equation}
f^{lJ}_{\nu_1,\nu_2}(\Omega_1,\Omega_{23}) = \sum_{m_J,m,m_l}
  g^{lJ}_{\nu_2,m_J}(\Omega_{23})
  \langle Jm_JJm|00\rangle \langle lm_l\frac{1}{2}\nu_1|Jm\rangle
   Y_{lm_l}(\bm{\hat{p}}_1),
\end{equation}
which can be written
\begin{equation}
f^{lJ}_{\nu_1\nu_2}(\Omega_1,\Omega_{23})=\sum_{m,m_l,m_l'}
  \frac{(-1)^{J+m}}{\sqrt{2J+1}}
  \langle l m_l \frac{1}{2} \nu_1 | Jm \rangle
  \langle l m_l' \frac{1}{2} \nu_2 | J-m \rangle
  Y_{lm_l}(\bm{\hat{p}}_1)Y_{lm_l'}(\bm{\hat{p}}_{23}).
\label{eq:f_spin_ang}
\end{equation}
The quantities in angled brackets are the Clebsch-Gordan coefficients and
$Y_{lm}$ are the spherical harmonics with
$\Omega_1$ representing the angles $(\theta_1,\phi_1)$
that describe the emission of neutron~1 with momentum
$\bm{p}_1=p_1\bm{\hat{p}}_1$ in the three-body
c.m. system and $\Omega_{23}$ representing $(\theta_{23},\phi_{23})$
that describe the the emission of the of neutron~2 in with momentum
$\bm{p}_{23}=p_{23}\bm{\hat{p}}_{23}$ the rest frame of the $2-3$ system.
Finally, our amplitude can be made antisymmetric under the exchange of
neutrons by adopting
\begin{equation}
\mathcal{M}_{\nu_1\nu_2} = \sum_c \left[ u_c(\tilde{E_1})
  f^c_{\nu_1\nu_2}(\Omega_1,\Omega_{23})
 - u_c(\tilde{E_2}) f^c_{\nu_2\nu_1}(\Omega_2,\Omega_{13}) \right],
\end{equation}
with $\Omega_2$, $\Omega_{13}$, and related quantities
defined analogously to the above.

Our approach only considers $n+\alpha$
configurations for the description of ${}^5{\rm He}$ states;
this approximation is well justified in this case because
the thresholds for other configurations, such as $d+\text{T}$,
are located much higher in excitation energy.
Below neutron energies of 20~MeV, it is found that considering
$l\le 3$ is sufficient and that one level plus a constant $R^\infty$
for each channel allows for a good fit to be obtained.
We consider here $l=0$ and~1 transitions involving the $1/2^+$, $1/2^-$, and
$3/2^-$ $n+\alpha$ partial waves.

The scattering of neutrons by $\alpha$~particles is very well studied and
$R$-matrix parameters are available~\cite{Sta72}.
The quality of fit to $n+\alpha$ scattering observables is comparable
to modern analyses (e.g., Ref.~\cite{Cso97}) that take into account
more multichannel data.
We utilize the $R$-matrix parameters given in
Table~2 of Ref.~\cite{Sta72}, with $R^\infty$
replaced by a background level at very high (1000 MeV) excitation energy,
which we refer to hereafter to as the $R^\infty$ state.
Both $l=1$ partial waves ($1/2^-$ and $3/2^-$) have a resonant states
at low excitation energy, in the range which can be populated by low-energy
$\text{T}+\text{T}$ reactions. Consequently, these partial waves are
expected to contribute significantly.
Feedings of both the important resonance state and the $R^\infty$ state
are considered. We choose the boundary condition constant for these
parameters so that the level shift vanishes for the
lowest-energy state in each partial wave.
The $l=0$ $1/2^+$ partial wave is non-resonant, but is included
due to its low angular momentum. We consider feeding of the $1/2^+$
50-MeV level
(which is in fact a background state), but not the $R^\infty$ state.
It should be pointed out that the $R^\infty$ state for this partial wave
is unphysical, as it has $\gamma_{c\lambda}^2< 0$.
Since this state contributes little to the phase shift in the region of
interest, this issue is not a concern for the present work.
The channel radius for the $n+\alpha$ parameters is 3.0~fm; for
$n+{}^5{\rm He}$ we have used 4.0~fm.

\subsection{Di-neutron emission}
\label{subsec:dineutron}

We consider here the emission of neutrons in a $l=0$ spin singlet state, with
the orbital angular momentum of the neutron pair with respect to
the $\alpha$-particle core also taken to be zero.
In this case we assume the amplitude for the process is given by
\begin{equation}
\mathcal{M}_{\nu_1\nu_2} = u_c(\tilde{E_3})
  (f^{0,1/2}_{\nu_1\nu_2}-f^{0,1/2}_{\nu_2\nu_1}),
\end{equation}
with $c=(0,1/2,nn)$ for di-neutron emission.
The energy dependence is described by the $R$-matrix expression
\begin{equation}
u_c(E_3)=\left[\frac{P_3 P_{12}}{p_3 \,p_{12}}\right]^{1/2}
  e^{i(\omega_3-\Phi_3)} e^{i(\omega_{12}-\Phi_{12})}
  \frac{\frac{A_c\gamma_c}{E_c-E_{12}}}
  {1-(S_{12}-B_c+iP_{12})R_c},
\label{eq:u_n_n}
\end{equation}
where the notation is analogous to that given in the previous section
and we also have assumed only a single level
such that $R_c=\gamma^2_c/(E_c-E_{12})$.
Note also that  the shift function vanishes for $l=0$ neutrons
and we take $B_c=0$ here.
Adopting $E_c=3.119$~MeV and $\gamma^2_c=31.95$~MeV, for a channel radius of
2.0~fm, reproduces the scattering length
and effective range of the Argonne~V18 potential~\cite{Wir95}
which are $-18.487$~fm and 2.840~fm, respectively.
In addition, the phase shifts below a neutron energy of 10~MeV are
reproduced to within 2.5~degrees with this choice.
The $(nn)+\alpha$ channel radius has been taken to be 3.5~fm.
The antisymmetric spin singlet state has been generated with
the aid of Eq.~(\ref{eq:f_spin_ang}), with $l=0$ and $J=1/2$.
Note that the there is no angular dependence in this case.

\subsection{Definition of particle energy spectra}

Considering both di-neutron emission and the sequential emission of neutrons
through ${}^5{\rm He}$ states, we arrive at the final form for our
matrix element:
\begin{equation}
\mathcal{M}_{\nu_1\nu_2} = \sum_c 
\left[u_c(12)
  f^{lJ}_{\nu_1\nu_2}(\Omega_1,\Omega_{23})
 - u_c(21) f^{lJ}_{\nu_2\nu_1}(\Omega_2,\Omega_{13})\right],
\end{equation}
where the sum is over three $n\alpha$ channels and one $nn$ channel.
The nature of the energy dependence of $u_c$ varies with the
channel type ($n\alpha$ versus $nn$); the $12$ and $21$ notation is used to
indicate direct $(12)$ and exchange $(21)$ terms.
In principle, all observables can now be calculated.
Our primary interest, however, is in calculating the particle energy spectra.
The particle distribution in the three-body c.m. system is given by
\begin{equation}
\frac{d^3N}{dE_i \,d\Omega_i \, d\Omega_j}=
\sum_{\nu_1,\,\nu_2}|\mathcal{M}_{\nu_1\nu_2}|^2\,p_ip_{jk}\mathcal{J}_{ijk},
\label{eq:particle_dist}
\end{equation}
where the product of factors $p_ip_{jk}\mathcal{J}_{ijk}$ is the three-body
phase space~\cite{Bro64,Ohl65}, and $\mathcal{J}_{ijk}$ is the Jacobian for the
transformation from the $(\tilde{E_i},\Omega_i,\Omega_{jk})$ system to
the $(E_i,\Omega_i,\Omega_j)$ system.

In order to extract the particle energy distributions, it is necessary to
integrate out the angular variables. This task can be accomplished most
easily by transforming to the $(\tilde{E_i},\Omega_i,\Omega_{jk})$ system:
\begin{eqnarray}
\frac{dN}{dE_i} &=& \frac{M}{m_j+m_k}\frac{dN}{d\tilde{E_i}} 
   = \frac{M}{m_j+m_k} \int d\Omega_i \, d\Omega_{jk}
   \frac{d^3N}{d\tilde{E_i} \, d\Omega_i \, d\Omega_{jk}} \\
  &=& \frac{M}{m_j+m_k} \int \frac{d\Omega_i \,
    d\Omega_{jk}}{\mathcal{J}_{ijk}}
    \frac{d^3N}{dE_i \, d\Omega_i \, d\Omega_j} \\
  &=&\frac{M}{m_j+m_k} \int d\Omega_i \, d\Omega_{jk}
 \,\,p_ip_{jk}\sum_{\nu_1,\,\nu_2}
  \left|M_{\nu_1\nu_2}\right|^2.
\end{eqnarray}

\subsection{Evaluation of spin and angle-dependent functions}

In order to proceed further, it is necessary to evaluate the square of
the matrix element, summed over spin projections.
In doing so, two types of sums arise:
\begin{eqnarray}
\label{eqn:dir}
W^{(i)}_{lJl'J'} &=& (4\pi)^2 \sum_{\nu_1,\,\nu_2}
  f^{lJ}_{\nu_1\nu_2}(\Omega_i,\Omega_{jk})
  f^{l'J'*}_{\nu_1\nu_2}(\Omega_i,\Omega_{jk}) \quad {\rm and} \\
\label{eqn:exch}
W^{(12)}_{lJl'J'} &=& (4\pi)^2 \sum_{\nu_1,\,\nu_2}
  f^{lJ}_{\nu_1\nu_2}(\Omega_1,\Omega_{23})
  f^{l'J'*}_{\nu_2\nu_1}(\Omega_2,\Omega_{13}),
\end{eqnarray}
where the factors of $(4\pi)^2$ have been inserted for later convenience.
The first type of term can be evaluated using standard techniques
(see e.g. Ref.~\cite{Dev57}):
\begin{eqnarray}
W^{(i)}_{lJl'J'} &=& [(2J+1)(2J'+1)]^{1/2}(2l+1)(2l'+1)
  \times\nonumber \\ && \quad
  \sum_k \langle l0l'0|k0\rangle^2 W^2(klJ'\frac{1}{2};l'J)
  (-1)^k P_k(\cos\gamma_{jk}),
\label{eq:wi}
\end{eqnarray}
where the $W$ without subscripts is the Racah coefficient,
$P_k$ is the Legendre polynomial of order $k$, and
\begin{equation}
  \cos\gamma_{jk} = \bm{\hat{p}}_i \cdot \bm{\hat{p}}_{jk}.
\end{equation}

The second type of term arises from antisymmetrization and is more
complicated to evaluate, due to the fact that, as written, it
depends on two sets of angular variables that are not independent.
The angular variables are represented by the unit vectors of the
momenta. By using Eqs.~(\ref{eq:momentum_1}) and~(\ref{eq:momentum_2}),
$\bm{\hat{p}}_{23}$ and $\bm{\hat{p}}_{13}$ can be eliminated using
\begin{eqnarray}
\bm{\hat{p}}_{23} &=& \frac{p_2}{p_{23}}\bm{\hat{p}}_2 +
                 \frac{p_1}{p_{23}}\frac{m_2}{m_2+m_3}\bm{\hat{p}}_1 \\
\bm{\hat{p}}_{13} &=& \frac{p_1}{p_{13}}\bm{\hat{p}}_1 +
                 \frac{p_2}{p_{13}}\frac{m_1}{m_1+m_3}\bm{\hat{p}}_2
\end{eqnarray}
so that the expression only depends on the angular variables
$\bm{\hat{p}}_1$ and $\bm{\hat{p}}_2$.
The spherical harmonics harmonics can then be evaluated for these
substitutions using the following addition theorem~\cite{Ros58}
\begin{equation}
c^l\,Y_{lm}(\bm{\hat{c}})=
  \sum_{\substack{\lambda_1+\lambda_2=l \\ \nu_1+\nu_2=m}}
  a^{\lambda_1}b^{\lambda_2}
  \langle \lambda_1\nu_1 \lambda_2 \nu_2|l m\rangle
  \sqrt{\frac{4\pi (2l+1)!}{(2\lambda_1+1)!(2\lambda_2+1)!}}
  Y_{\lambda_1\nu_1}(\bm{\hat{a}})
  Y_{\lambda_2\nu_2}(\bm{\hat{b}}),
\end{equation}
where $\bm{c}=\bm{a}+\bm{b}$ with
$\bm{a}=a\bm{\hat{a}}$, $\bm{b}=b\bm{\hat{b}}$, and
$\bm{c}=c\bm{\hat{c}}$.
The second type of term is then found to be:
\begin{eqnarray}
W^{(12)}_{lJl'J'} &=&
  (-1)^{J+J'}[(2J+1)(2J'+1)(2l+1)!(2l'+1)!]^{1/2}
  (2l+1)(2l'+1)\times\nonumber \\
  &&
  \sum_{\substack{\lambda_1+\lambda_1'=l \\ \lambda_2+\lambda_2'=l' \\
    \lambda_3, \lambda_3', \lambda_3'', k}}
  \left(\frac{p_2}{p_{23}}\right)^{\lambda_1}
  \left(\frac{m_2}{m_2+m_3}\frac{p_1}{p_{23}}\right)^{\lambda_1'}
  \left(\frac{p_1}{p_{13}}\right)^{\lambda_2}
  \left(\frac{m_1}{m_1+m_3}\frac{p_2}{p_{13}}\right)^{\lambda_2'}
  \times\nonumber \\ &&
  \left[\frac{(2\lambda_3+1)(2\lambda_3'+1)}{(2\lambda_1)!(2\lambda_1')!
    (2\lambda_2)!(2\lambda_2')!}\right]^{1/2} (2\lambda_3''+1)
  \times\nonumber \\ &&
  \langle\lambda_1  0\lambda_2' 0|\lambda_3 0\rangle
  \langle\lambda_1' 0\lambda_2  0|\lambda_3' 0\rangle
  \langle l         0\lambda_3' 0|k 0\rangle
  \langle l'        0\lambda_3  0|k 0\rangle
  \times\nonumber \\ &&
  \left\{ \begin{array}{lll} \lambda_1  & \lambda_2' & \lambda_3  \\
                             \lambda_1' & \lambda_2  & \lambda_3' \\
                             \l         & l'         & \lambda_3''
  \end{array}\right\}
  \left\{ \begin{array}{lll} \frac{1}{2} & J           & l \\
                             J'          & \frac{1}{2} & l' \\
                             l'          & l           & \lambda_3''
  \end{array}\right\}
  W(l' l\lambda_3\lambda_3';\lambda_3''k)
  (-1)^{\lambda_3'+\lambda_3''-l} P_k(\cos\delta_{12}),
\label{eq:w12}
\end{eqnarray}
where $\{\}$ indicates the Wigner 9-$J$ symbol and
\begin{equation}
  \cos\delta_{12} = \bm{\hat{p}}_1 \cdot \bm{\hat{p}}_2.
\end{equation}

The functions $W^{(i)}_{lJl'J'}$ are real and are invariant under the
interchange of $(l,J)$ and $(l',J')$, while $W^{(12)}_{lJl'J'}$
are also real and are invariant under the interchange of $(l,J)$
and $(l',J')$ and particle labels~1 and~2.
These functions are tabulated in Table~\ref{tab:w} for the partial wave
combinations considered here.

\begingroup
\squeezetable
\begin{table*}[tb]
\caption{\label{tab:w}The angular functions $W^{(i)}_{lJl'J'}$ given by
Eq.~(\protect\ref{eq:wi}) and $W^{(12)}_{lJl'J'}$ given by
Eq.~(\protect\ref{eq:w12}), for the partial wave combinations considered here.}
\begin{ruledtabular}
\begin{tabular}{llllll}
$l$ & $J$ & $l'$ & $J'$ & $W^{(i)}_{lJl'J'}$ &
  $W^{(12)}_{lJl'J'}$ \\[1pt] \hline
0 & $\frac{1}{2}$ & 0 & $\frac{1}{2}$ & 1 & $-1$ \\
0 & $\frac{1}{2}$ & 1 & $\frac{1}{2}$ & $-\cos\gamma_{jk}$ &
  $\frac{m_1}{m_1+m_3} \frac{p_2}{p_{13}}+
  \frac{p_1}{p_{13}}\cos\delta_{12}$ \\
0 & $\frac{1}{2}$ & 1 & $\frac{3}{2}$ & $-\sqrt{2}\cos\gamma_{jk}$ &
  $\sqrt{2}\left( \frac{m_1}{m_1+m_3} \frac{p_2}{p_{13}}+
  \frac{p_1}{p_{13}}\cos\delta_{12} \right)$ \\
1 & $\frac{1}{2}$ & 1 & $\frac{1}{2}$ & 1 &
  $-\frac{p_1p_2}{p_{13}p_{23}}\left[ \left(
  \frac{p_2}{p_1}\frac{m_1}{m_1+m_3}+\frac{p_1}{p_2}\frac{m_2}{m_2+m_3}\right)
  \cos\delta_{12} + 1 +
  \frac{m_1m_2}{(m_1+m_3)(m_2+m_3)}\right]$ \\
1 & $\frac{1}{2}$ & 1 & $\frac{3}{2}$ & $\sqrt{2}P_2(\cos\gamma_{jk})$ &
  $-\sqrt{2}\frac{p_1p_2}{p_{13}p_{23}}\left[ \left(
  \frac{p_2}{p_1}\frac{m_1}{m_1+m_3}+\frac{p_1}{p_2}\frac{m_2}{m_2+m_3}\right)
  \cos\delta_{12} + P_2(\cos\delta_{12}) +
  \frac{m_1m_2}{(m_1+m_3)(m_2+m_3)}\right]$ \\
1 & $\frac{3}{2}$ & 1 & $\frac{3}{2}$ & $1+P_2(\cos\gamma_{jk})$ &
 $-\frac{p_1p_2}{p_{13}p_{23}}\left[2\left(
  \frac{p_2}{p_1}\frac{m_1}{m_1+m_3}+\frac{p_1}{p_2}\frac{m_2}{m_2+m_3}\right)
  \cos\delta_{12} + 1 + P_2(\cos\delta_{12})
  +\frac{2m_1m_2}{(m_1+m_3)(m_2+m_3)}\right]$
\end{tabular}
\end{ruledtabular}
\end{table*}
\endgroup

Due to our assumption of a $J=0$ initial state, the particles are
emitted isotropically; these functions thus describe the angular
correlations between the particles. Note that the particle distribution
is a function of two variables, which can be taken to be $E_i$
and $\cos\gamma_{jk}$. From these two quantities, all other needed
energies, momentum magnitudes, and angles
can be calculated from the kinematics relationships.

\subsection{Calculation of particle energy spectra}
\label{subsec:calc_spec}

We can now write
\begin{equation}
(4\pi)^2\sum_{\nu_1,\,\nu_2}|\mathcal{M}_{\nu_1\nu_2}|^2 =\sum_{c,c'}
g^{(1)}_{cc'}+g^{(2)}_{cc'}+g^{(12)}_{cc'},
\end{equation}
where
\begin{eqnarray}
g^{(1)}_{cc'} &=& u_c(12) \, u_{c'}^*(12) \,
  W^{(1)}_{lJl'J'}(\cos\gamma_{23}) \\
g^{(2)}_{cc'} &=& u_c(21) \, u_{c'}^*(21) \,
  W^{(2)}_{lJl'J'}(\cos\gamma_{13}) \\
g^{(12)}_{cc'} &=& -2 \, {\rm Re}[\, u_c(12) \, u_{c'}^*(21) \,] \,
  W^{(12)}_{lJl'J'}.
\end{eqnarray}
If the neutrons were distinguishable, the $g^{(1)}$ contribution would
be the neutron~1 distribution and $g^{(2)}$ would be the
neutron~2 distribution. The $g^{(12)}$ term arises
from treating the neutron as indistinguishable fermions.
In the case of neutron emission via intermediate ${}^5{\rm He}$ states, we
take neutron~1 to the first neutron emitted and neutron~2 to be the second.
The neutron energy distribution can be calculated:
\begin{eqnarray}
\frac{dN}{dE_i} &=& \frac{M}{m_j+m_k}\frac{1}{2}
  \int_{-1}^1 d(\cos\gamma_{jk})\,p_ip_{jk}\,\left[ \sum_{c,c'}
  g^{(1)}_{cc'}+g^{(2)}_{cc'}+g^{(12)}_{cc'}\right] \label{eq:dnde} \\
&=& \frac{dN^{(1)}}{dE_i} + \frac{dN^{(2)}}{dE_i} +
    \frac{dN^{(12)}}{dE_i}.
\end{eqnarray}
The first term will be called the primary distribution, the second
the secondary distribution, and the third the exchange distribution.

If only $n\alpha$ channels are present, the calculation of the
primary contribution can be simplified, because $u_c$ then only depends on
$\tilde{E_1}$, and $\tilde{E_1}$, $p_1$, and $p_{23}$ are independent
of $\cos\gamma_{23}$. The result is:
\begin{equation}
\frac{dN^{(1)}}{dE_1}=\frac{M}{m_2+m_3}p_1p_{23}
\sum_c \left|u_c(\tilde{E_1})\right|^2,
\end{equation}
which is free from any angular correlation or interference effects.
Note also that it is in this situation that the $R$-matrix energy
distribution formula given in Ref.~\cite{Bar88} is recovered.
For the $\alpha$-particle energy distribution, all of the contributions
must be calculated by numerical integration, but this task is simplified
by noting that
\begin{equation}
\frac{dN^{(1)}}{dE_3}=\frac{dN^{(2)}}{dE_3}.
\end{equation}

If only the $nn$ (di-neutron) channel is present, the calculation also
simplifies. In this case we have
\begin{equation}
\frac{dN^{(1)}}{dE_i}=\frac{dN^{(2)}}{dE_i}=
\frac{1}{2}\frac{dN^{(12)}}{dE_i}.
\label{eq:di_neutron_contrib}
\end{equation}
For the neutron energy distribution, these must be evaluated by numerical
integration, but for the $\alpha$-particle energy distribution
we have
\begin{equation}
\frac{dN}{dE_3}=\frac{4M}{m_1+m_2}p_3p_{12}\left|u_c(\tilde{E_3})\right|^2,
\end{equation}
which is also in the form given by Ref.~\cite{Bar88}.

In the general case, all three contributions to the distribution must
be calculated using numerical integration. We do note that
the overall contributions of the primary and secondary distributions
are equal, i.e., that
\begin{equation}
\int_0^{\frac{m_j+m_k}{M}E_{tot}} \frac{dN^{(1)}}{dE_i} \, dE_i =
\int_0^{\frac{m_j+m_k}{M}E_{tot}} \frac{dN^{(2)}}{dE_i} \, dE_i.
\end{equation}

\subsubsection{Spectra for channels in isolation}
\label{subsubsec:isolation}

\begin{figure}[p]
\includegraphics[width=0.6\textwidth]{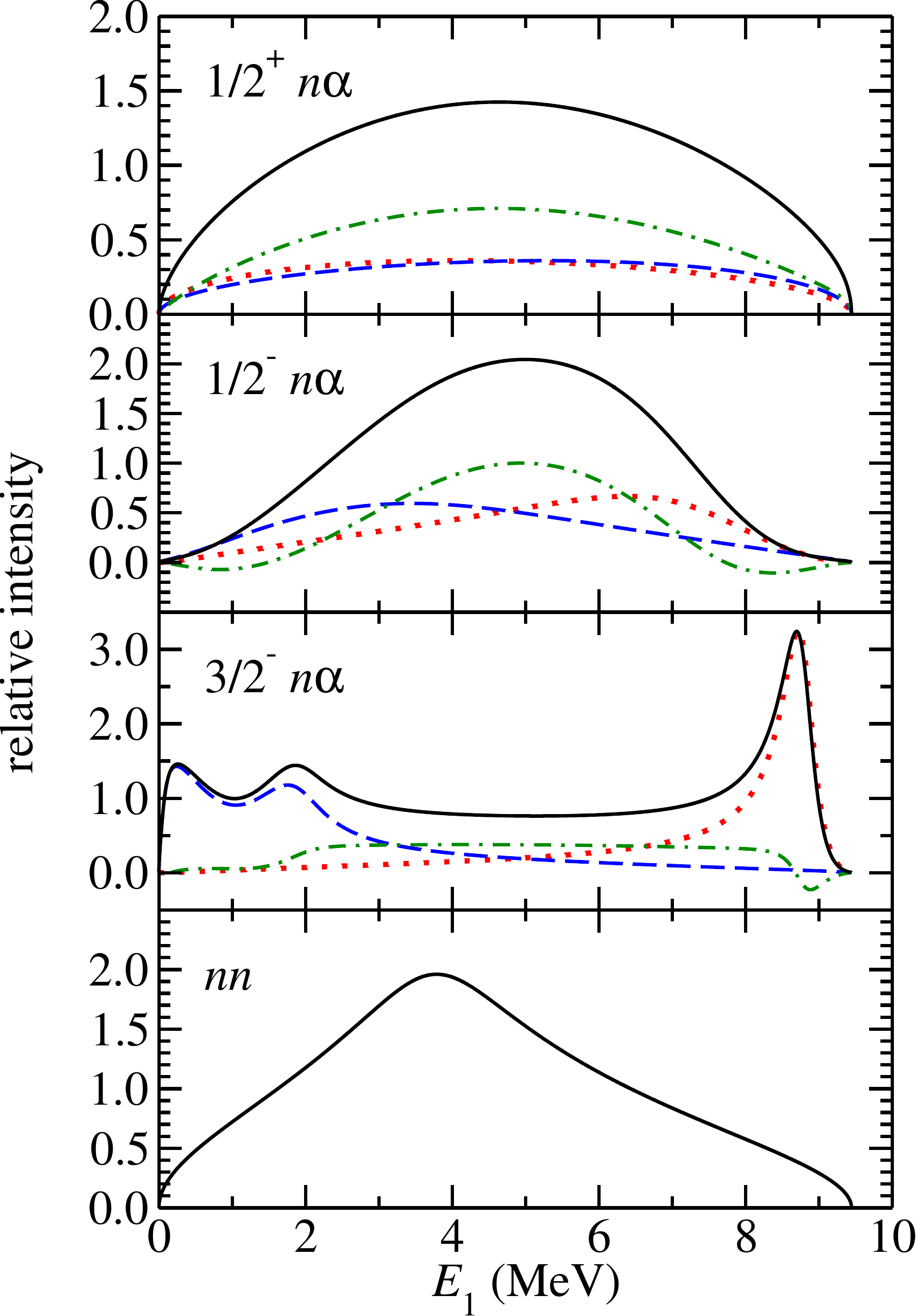}
\caption{\label{fig:ndist_chan}Neutron energy distributions for
each channel considered separately. The primary, secondary, exchange,
and total are given by the dotted, dashed, dot-dashed, and solid curves,
respectively. Only the total is shown for the $nn$ case.}
\end{figure}

\begin{figure}[p]
\includegraphics[width=0.6\textwidth]{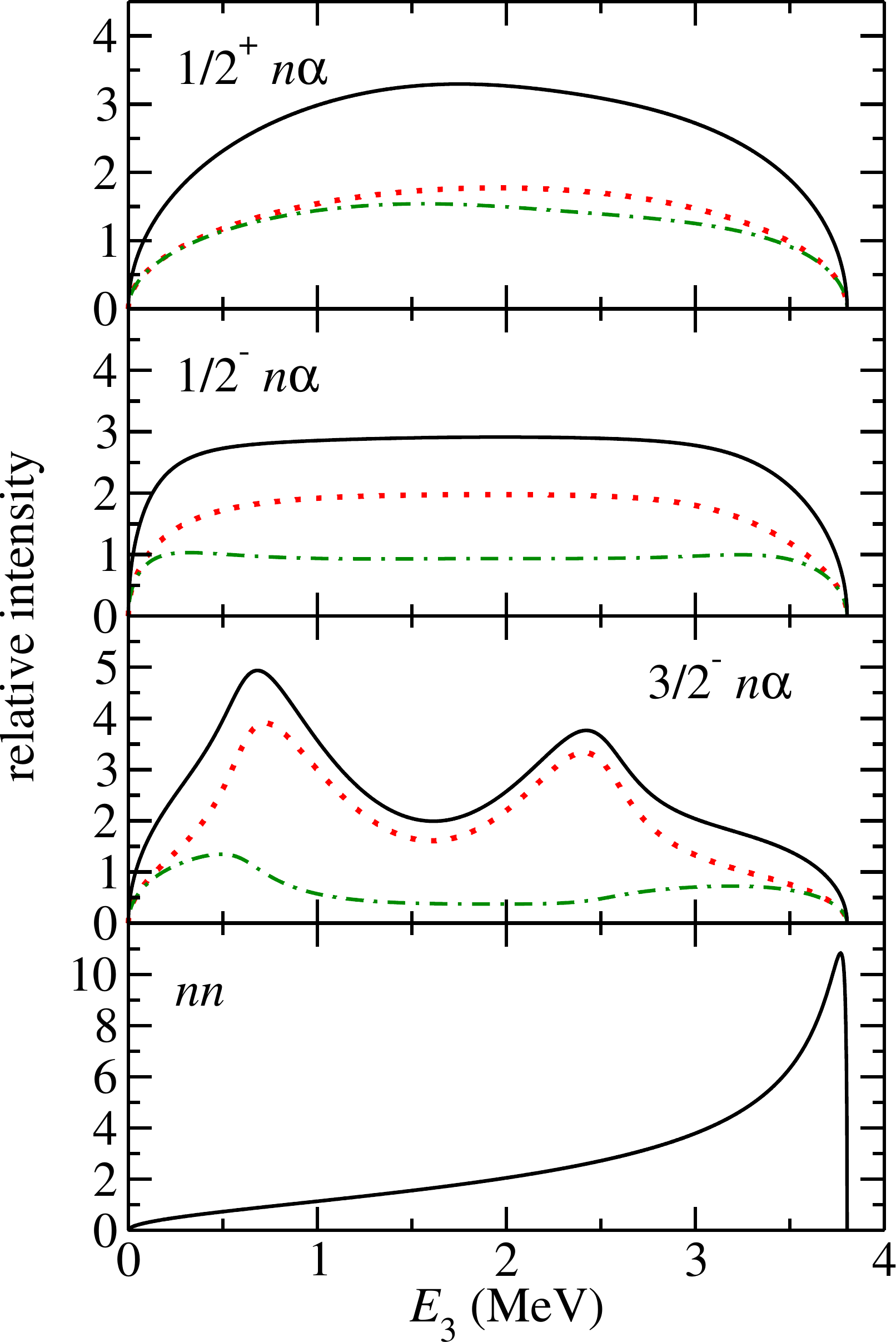}
\caption{\label{fig:adist_chan}Alpha-particle energy distributions for
each channel considered separately. The primary plus secondary, exchange,
and total are given by the dotted, dot-dashed, and solid curves, respectively.
Only the total is shown for the $nn$ case.}
\end{figure}

We will next investigate the nature of the particle energy distributions
resulting for each channel in isolation. 
Note that we make no effort in this section to adjust the feeding
factor parameters of the model to fit experimental data; this is
done below in Sec.~\ref{sec:data}.
For the $l=1$ $n\alpha$
channels, we have taken the background feeding to be zero.
Each channel thus has only a single feeding factor, which has been adjusted
so that $\int \frac{dN}{dE_i}{\scriptstyle dE_i}=10$.
The results are shown in Fig.~\ref{fig:ndist_chan} for the neutron
energy distributions and in Fig.~\ref{fig:adist_chan} for the
$\alpha$-particle energy distributions.
For the $n\alpha$ channels, the primary, secondary, exchange, and total
contributions are shown for neutron energy distributions, and
the primary plus secondary, exchange, and total contributions are
shown for the $\alpha$-particle distributions.
For the $nn$ channel, only the total is shown, since, as shown by
Eq.~(\ref{eq:di_neutron_contrib}), the sub-contributions are all proportional.
It is interesting to note that the interference introduced by
antisymmetrization has a general tendency to be constructive
in all cases investigated. This point is discussed further below in
Sec.~\ref{sec:heavy}.

\paragraph*{$1/2^+$ $n\alpha$:}

The particle spectra for this channel are rather featureless.
Both the neutron and $\alpha$-particle spectra closely approximate
elliptical energy distributions, characteristic of uniform
phase space population.

\paragraph*{$1/2^-$ $n\alpha$:}

The first excited state of ${}^5{\rm He}$ gives rise to a broad peak
in the primary neutron spectrum, while the secondary neutron spectrum
is also broad, but peaks at a lower neutron energy.
The effect of antisymmetrization is to make the overall spectrum
narrower, with relatively little strength near the endpoints of the spectrum.
The $\alpha$-particle spectrum for this channel is relatively flat,
except near the endpoints.

\paragraph*{$3/2^-$ $n\alpha$:}

The ground state of ${}^5{\rm He}$ gives rise to a narrow peak in the
primary neutron spectrum near the maximum neutron energy.
The secondary neutron spectrum shows a double-peaked feature below 2~MeV.
This structure results from the
$W^{(i)}=1+P_2(\cos\gamma_{jk})$ angular correlation between the
primary and secondary neutrons, which implies a strong tendency for
the neutrons to be emitted in the same or opposite directions, but
not perpendicular to each other.
Due to the recoil of the ${}^5{\rm He}$ intermediate state, this correlation
affects the secondary neutron energy distribution.
This angular correlation also gives rise to a double peak in the
$\alpha$-particle energy spectrum.
These effects on the particle energy spectra due to angular correlations were
understood over 50 years ago~\cite{Bam57,Jar58}, and were observed for the
the $\alpha$-particle energy spectrum
at higher $\text{T}+\text{T}$ energies~\cite{Jar58}.
Due to the relatively small energy overlap between primary and secondary
spectra, the effect of antisymmetrization on the overall spectra is
less important for this channel.

\paragraph*{$nn$:}

In this case, the neutron energy spectrum peaks just below 4~MeV and has
considerably less strength near the endpoints compared to the
$1/2^+$~$n\alpha$ channel which has the same quantum numbers.
The $\alpha$-particle spectrum has a very distinctive peak near the
maximum energy that is associated with the two neutrons being emitted in
nearly the same direction with a low relative energy.
Similar results for the effect of the $nn$ interaction on the
$\alpha$-particle spectrum were found in the calculations of
Lacina, Ingley, and Dorn~\cite{Lac65}.

\subsubsection{Interference between channels}

Another way to decompose Eq.~(\ref{eq:dnde}) that is useful when
considering multiple channels is
\begin{equation}
\frac{dN}{dE_i} = \sum_c \frac{dN_{cc}}{dE_i} +
  \sum_{\substack{c,c' \\c\neq c'}} \frac{dN_{cc'}}{dE_i},
\label{eq:chan_int}
\end{equation}
where the second sum is due to interference effects between channels and
\begin{equation}
\frac{dN_{cc'}}{dE_i} = \frac{M}{m_j+m_k}\frac{1}{2}
  \int_{-1}^1 d(\cos\gamma_{jk})\,p_ip_{jk}\,\left[
  g^{(1)}_{cc'}+g^{(2)}_{cc'}+g^{(12)}_{cc'}\right].
\end{equation}

\begin{figure}[p]
\includegraphics[width=0.6\textwidth]{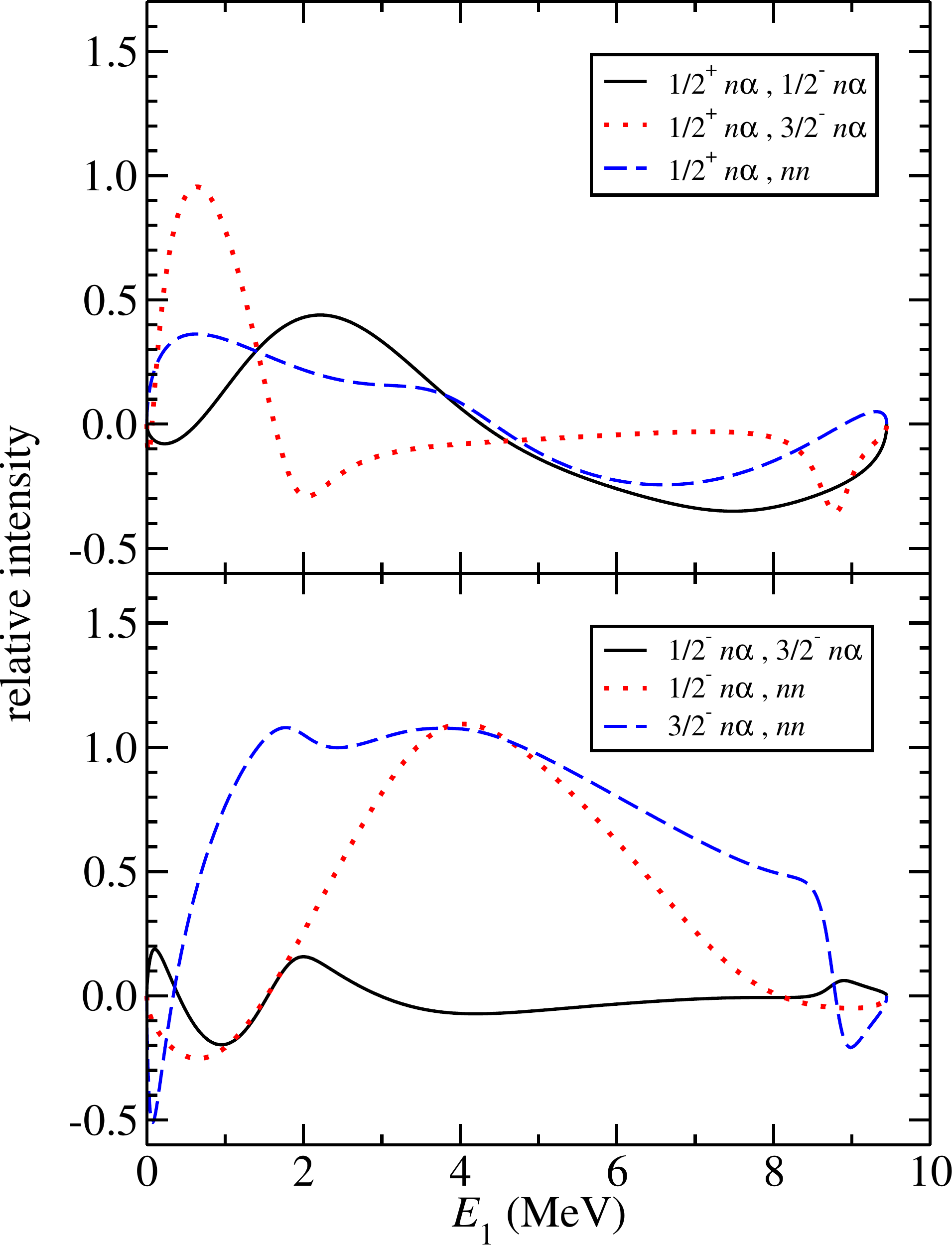}
\caption{\label{fig:n_interference}Interference contributions to the
neutron energy distributions for partial wave combinations indicated.}
\end{figure}

\begin{figure}[p]
\includegraphics[width=0.6\textwidth]{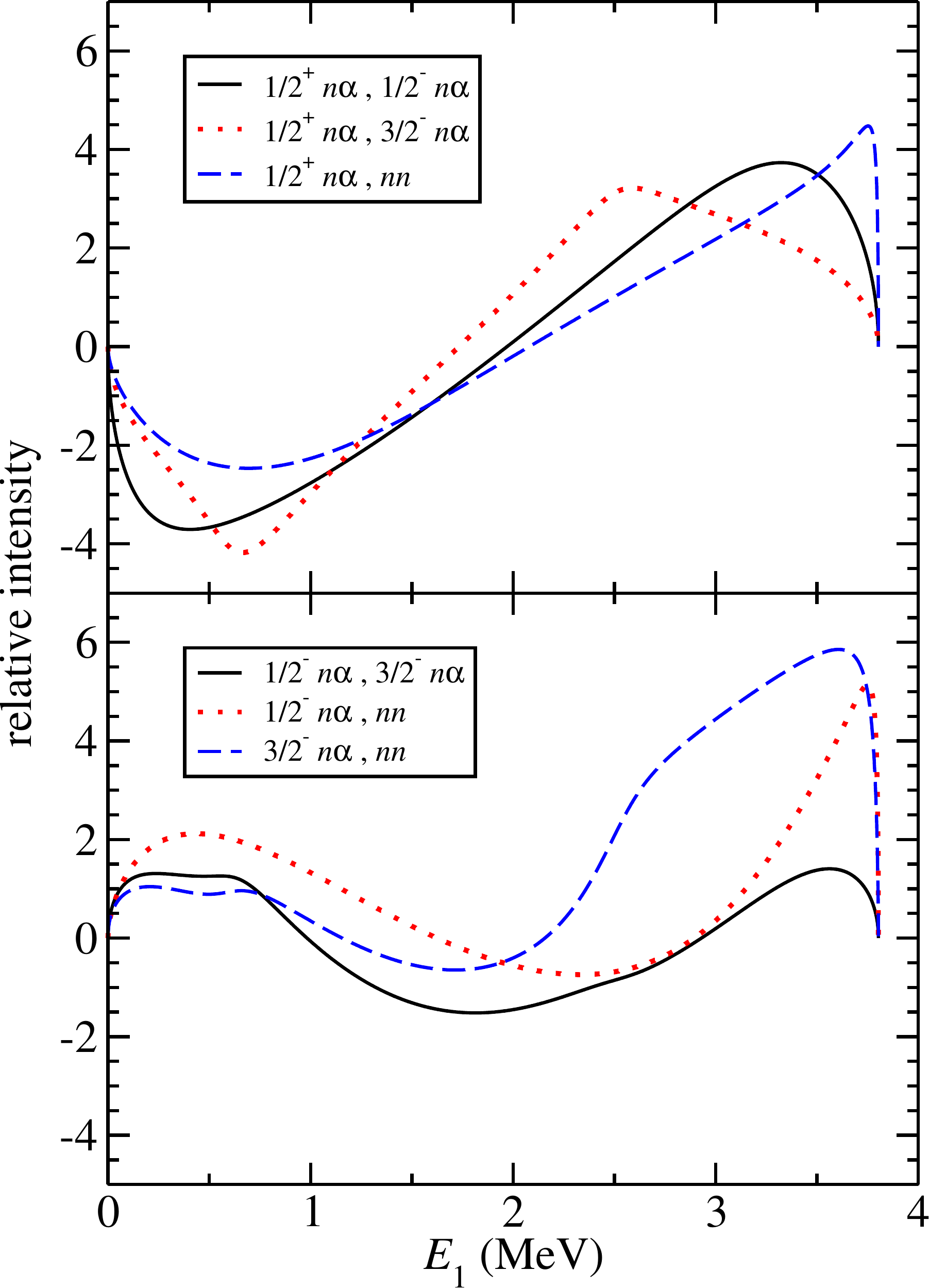}
\caption{\label{fig:a_interference}Interference contributions to the
$\alpha$-particle energy distributions for partial wave combinations
indicated.}
\end{figure}

The channel interference contributions to the particle spectra are shown
in Figs.~\ref{fig:n_interference} and~\ref{fig:a_interference},
for the partial wave combinations under consideration.
The same feeding factors were used as for the calculations shown
in Figs.~\ref{fig:ndist_chan} and~\ref{fig:adist_chan}.
Note that the signs of the interference contributions are determined
by the relative signs of the feeding factors.
The effects are seen to be substantial, comparable in magnitude to
the single-channel contributions.
In addition, note that the contribution of these effects, integrated
over $E_i$, does not vanish.

\section{Fits and comparisons to experimental data}
\label{sec:data}

In this section, fits and comparisons to experimental neutron and
$\alpha$-particle spectra from low-energy $\text{T}+\text{T}$ reactions
(below $E_{c.m.}=100$~keV) are presented.

\subsection{Neutron spectrum}

\begin{figure}[p]
\includegraphics[width=0.6\textwidth]{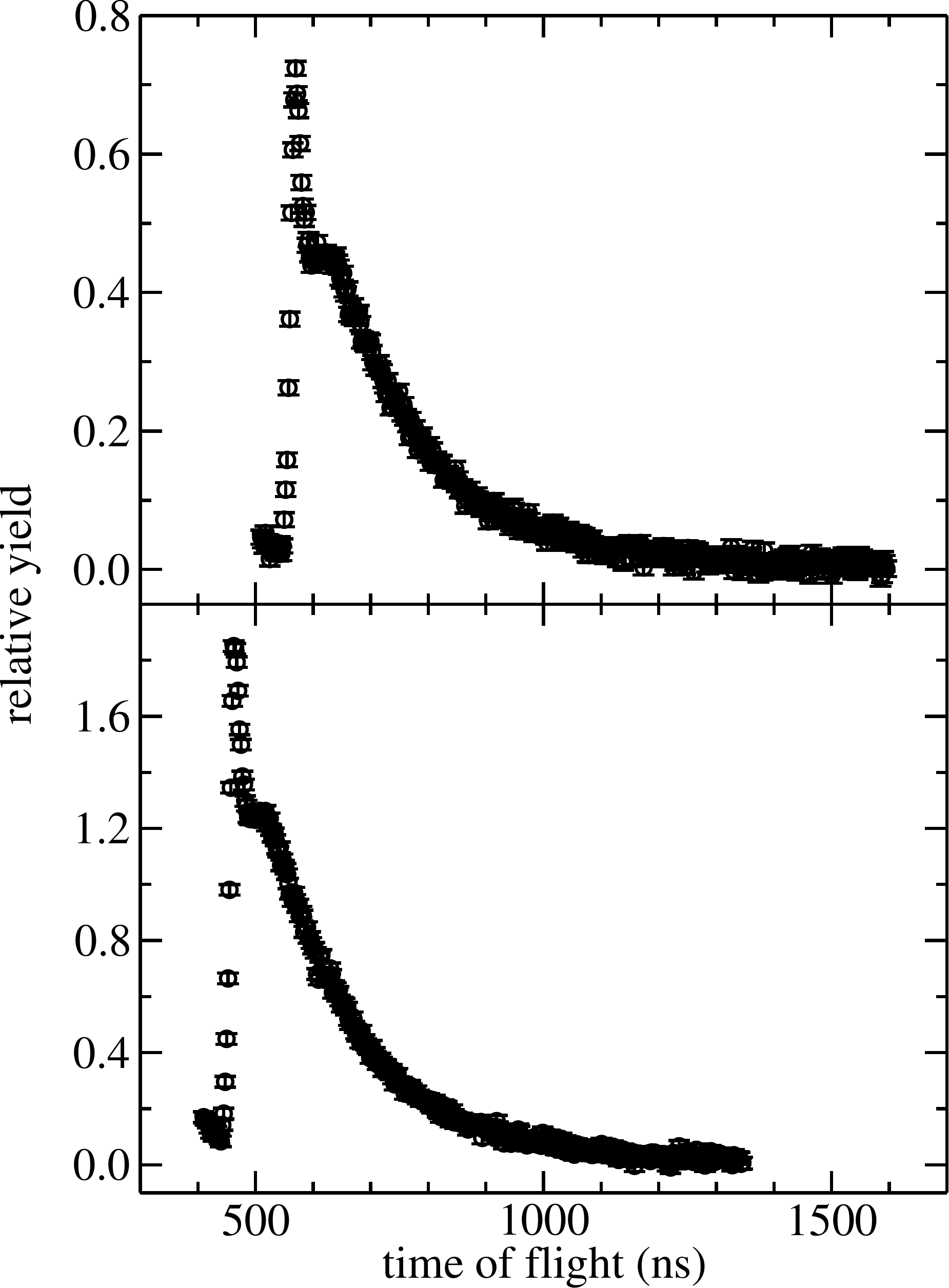}
\caption{\label{fig:time_plot}The neutron time-of-flight spectra from
the 22.2-m detector (top) and 20.1-m detector (bottom) used for fitting.}
\end{figure}

The $\text{T}(t,2n)$ neutron spectrum at an effective $E_{c.m.}$ of 16~keV has
recently been measured at the NIF~\cite{Say13}.
The neutrons were detected in two liquid scintillators along separate lines
of sight located 20.1 and 22.2~m from the source, respectively.
The experiment provides raw data in the form of digitized
currents from the detectors versus time. The data from the 22.2-m detector
has been presented in Fig.~2 of Ref.~\cite{Say13}, where the time of flight has
been converted to a nominal neutron energy and the points have been rebinned.
Here we will present additional fits to these data.
The fits to the raw data utilize the $R$-matrix
description of the neutron spectrum for $E_{c.m.}=16$~keV, and
take into account the following effects~\cite{Say13}: thermal broadening due
to the Maxwell-Boltzmann distribution of particle velocities in the plasma,
neutron attenuation and scattering between the $\text{T}(t,2n)$ reaction
source and the detector, the light output response of each detector,
and the time response of each detector.
Finally, the background from the $T(d,n)$ reaction, which was measured
separately, was added to the model spectra.
The data fitted here are identical to those reported in Ref.~\cite{Say13},
except that some additional points at longer times of flight have
been included (giving 812 data points in total),
and the errors on the data have been increased by assuming that the
attenuation correction has a 20\% uncertainty and that the scattering
correction has a 50\% uncertainty. The raw time-of-flight data that are fitted
are shown in Fig.~\ref{fig:time_plot}; note the narrow peak near 500~ns
is the peak in the neutron spectrum from the ${}^5{\rm He}$ ground state.

\begin{table}[p]
\caption{The $\chi^2_{\rm min}$ values obtained for the various feeding factor
assumptions. The presence of a \checkmark\ indicates that a particular feeding
factor was varied in the fit; a total of 812 data points were fitted.}
\label{tab:fit_results}
\begin{tabular}{c|c|cc|cc|c|c} \hline\hline
fit no. & \multicolumn{5}{c|}{$n\alpha$} & $nn$ & $\chi^2_{\rm min}$ \\ \hline
& $1/2^+$ & \multicolumn{2}{c|}{$1/2^-$} & \multicolumn{2}{c|}{$3/2^-$} & & \\
& $A$        & $A_1$      & $A_2$      & $A_1$      & $A_2$      & $A$        &   \\ \hline
1  &            & \checkmark &            & \checkmark &            &            & 2165 \\
2  & \checkmark & \checkmark &            & \checkmark &            &            & 1316 \\
3  &            & \checkmark &            & \checkmark &            & \checkmark & 1309 \\
4  &            & \checkmark &            & \checkmark & \checkmark &            & 1285 \\
5  &            & \checkmark & \checkmark & \checkmark &            &            & 1095 \\
6  & \checkmark & \checkmark &            & \checkmark &            & \checkmark & 867 \\
7  & \checkmark & \checkmark &            & \checkmark & \checkmark &            & 996 \\
8  & \checkmark & \checkmark & \checkmark & \checkmark &            &            & 660 \\
9  &            & \checkmark & \checkmark & \checkmark & \checkmark &            & 1085 \\
10 &            & \checkmark &            & \checkmark & \checkmark & \checkmark & 920 \\
11 &            & \checkmark & \checkmark & \checkmark &            & \checkmark & 1162 \\
12 & \checkmark & \checkmark & \checkmark & \checkmark & \checkmark &            & 659 \\
13 & \checkmark & \checkmark &            & \checkmark & \checkmark & \checkmark & 850 \\
14 & \checkmark & \checkmark & \checkmark & \checkmark &            & \checkmark & 660 \\
15 &            & \checkmark & \checkmark & \checkmark & \checkmark & \checkmark & 667 \\
16 & \checkmark & \checkmark & \checkmark & \checkmark & \checkmark & \checkmark & 632 \\ \hline\hline
\end{tabular}
\end{table}

The spectra have been fitted using the $R$-matrix formalism described
above, assuming various combinations of the
four channels discussed in Sec.~\ref{subsubsec:isolation}.
In addition, we have considered non-zero background feeding factors for the
$1/2^-$ and $3/2^-$ $n\alpha$ channels, leading to a total of up 6 variable
parameters. Note that the feeding of the low-lying $1/2^-$ and/or
$3/2^-$ states in $n\alpha$ channels were always fitted and that if a feeding
factor was not fitted, its value has been assumed to be zero.
The $\chi^2_{\rm min}$ values obtained for the various feeding factor
assumptions are presented in Table~\ref{tab:fit_results}.

\begin{figure}[p]
\includegraphics[width=0.6\textwidth]{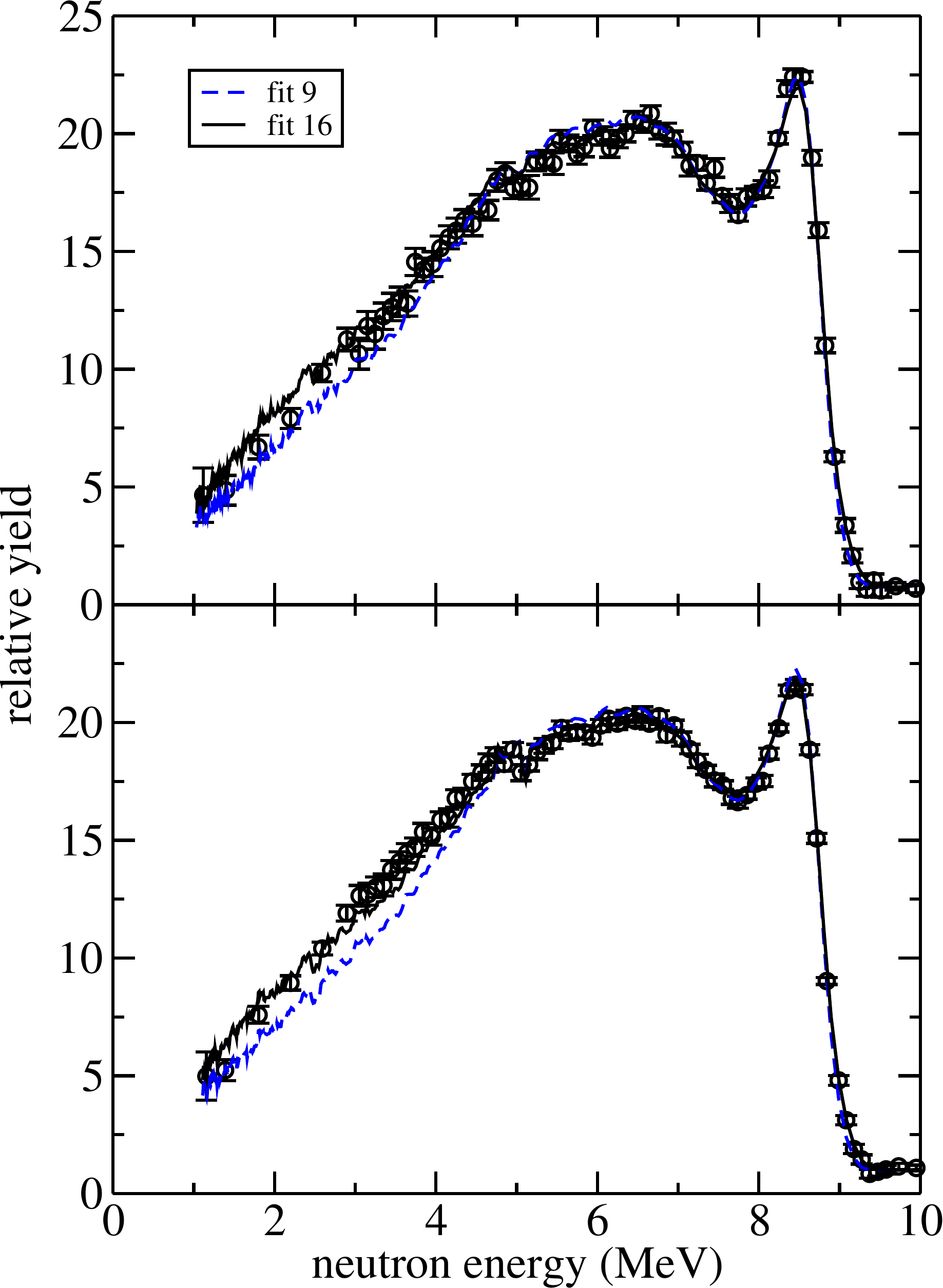}
\caption{\label{fig:sample_fits}Two fits to the raw neutron spectra,
plotted as a function of the nominal neutron energy, for
the 22.2-m detector (top) and 20.1-m detector (bottom).
It should be noted that these data, as well as those shown in
Fig.~\ref{fig:time_plot}, still include various experimental factors
such as thermal broadening, neutron light output, and detector time response.}
\end{figure}

Not surprisingly, the $\chi^2_{\rm min}$ decreases steadily as the number
of free parameters is increased. In Ref.~\cite{Say13}, only the $1/2^-$
and $3/2^-$ $n\alpha$ channels were considered. The fit presented there
is nearly identical to fit number~9 presented here, with the very small
changes arising from the changes in the data set discussed above.
The additional channels are seen to make a substantial increase in the
quality of the fit. However, we are cautious about placing a large emphasis
on this improvement, as the neutron spectrum data contain
neutron-energy-dependent systematic errors from the scattering and
attenuation corrections that may be comparable to this improvement
in fit (i.e., from $\chi^2_{\rm min}=1085$ to 632).
Two of the fits, numbers 9 and 16, are shown if Fig.~\ref{fig:sample_fits},
where the data have been re-binned and plotted versus the nominal
neutron energy.

\begin{figure}[p]
\includegraphics[width=0.6\textwidth]{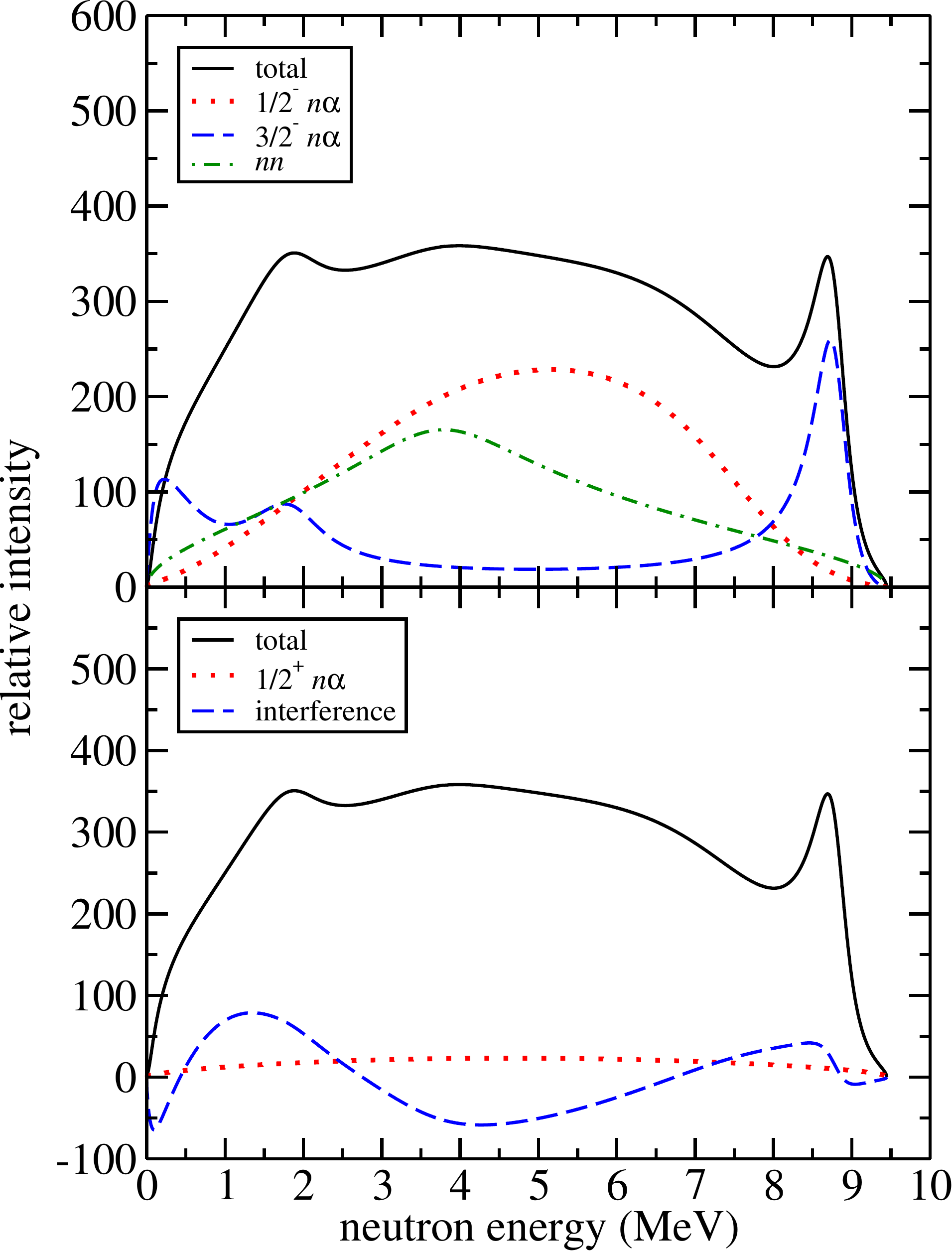}
\caption{\label{fig:decomp}The decomposition of fit~16 into its various
channel components and the net interference contribution.}
\end{figure}

The $R$-matrix parameters for fit~16 are shown in Table~\ref{tab:fit_params}.
Note that the uncertainties on the feeding factors are computed
assuming that uncertainties on the data are random and normally
distributed; as explained above this is not strictly the case and the
true uncertainties are larger (this is also why some fits are able achieve
a $\chi^2$ value which is less than the number of data points).
The decomposition of this fit into its various channel components and
the net interference contribution, according to Eq.~(\ref{eq:chan_int}),
is shown in Fig.~\ref{fig:decomp}.
It is seen that the fitted $1/2^-$ and $3/2^-$ $n\alpha$ channels,
and di-neutron channel, are substantial, with the $1/2^+$ $n\alpha$
channel contributing to a lesser degree. The net interference between
channels is also non-negligible.
In Ref.~\cite{Say13}, a branching ratio for the  the $1/2^-$
and $3/2^-$ $n\alpha$ channels was given, which is possible to do if
only these two channels are considered, since their interference term
is very small. In the general case, it is not possible to determine branching
ratios, due to the substantial interference contribution.

\begin{table}[p]
\caption{The $R$-matrix parameters for fit~16. Note that the boundary
condition parameter is $B=S(E_{c1})$ for the $n\alpha$ channels, and
$B=0$ for the $nn$ channel.
The $\gamma_{c\lambda}$ are defined to be the positive square roots of
$\gamma^2_{c\lambda}$ and the channel radii are given in
Sec.~\ref{sec:rmatrix}.}
\label{tab:fit_params}
\begin{tabular}{ccccc} \hline\hline
channel & $\lambda$ & $E_{c\lambda}$ & $\gamma_{c\lambda}^2$ &
  $A_{c\lambda}$ \\
        &           & (MeV)          & (MeV)                 & \\ \hline
$1/2^+$ $n\alpha$ & 1 & 50.00 & 12.00 & -18(3)   \\
$1/2^+$ $n\alpha$ & 2 & 1000  & -40   & 0        \\ \hline
$1/2^-$ $n\alpha$ & 1 & 6.43  & 12.30 & -18.2(3) \\
$1/2^-$ $n\alpha$ & 2 & 1000  & 300   & -306(16) \\ \hline
$3/2^-$ $n\alpha$ & 1 & 0.97  & 7.55  & 9.86(6)  \\
$3/2^-$ $n\alpha$ & 2 & 1000  & 300   & 155(9)   \\ \hline
$nn$              & 1 & 3.119 & 31.95 & 12.5(5)  \\ \hline\hline
\end{tabular}
\end{table}

\begin{figure}[p]
\includegraphics[width=0.6\textwidth]{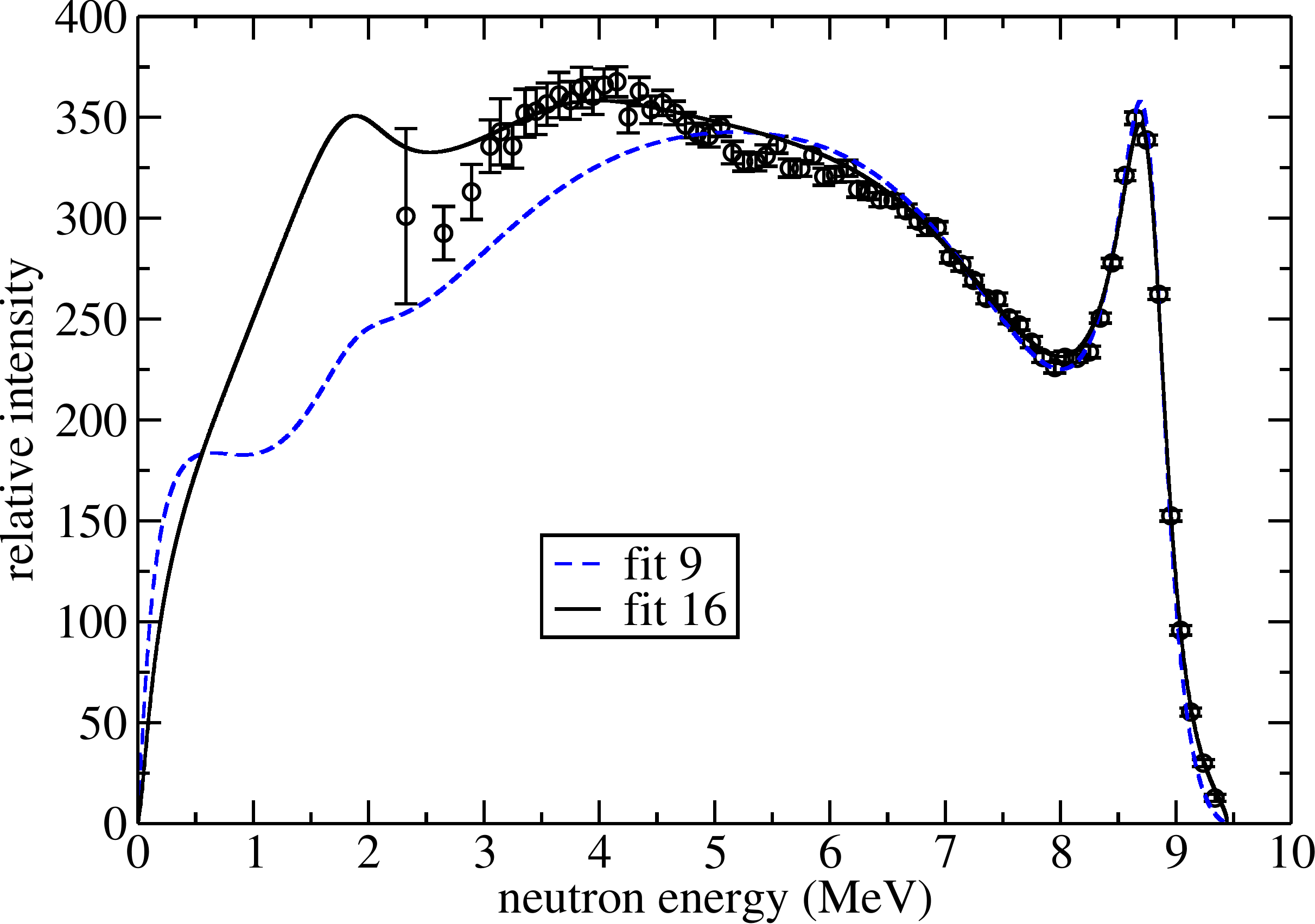}
\caption{\label{fig:decon}The deconvoluted neutron energy spectrum (points)
along with fits 9 and~16 (curves).}
\end{figure}

In order to facilitate future comparisons with other experiments and
calculations, it is desirable to present the neutron spectra in a
deconvoluted form, i.e., with the various efficiency, resolution, and
background corrections removed. General methods and considerations
for the deconvolution of nuclear science measurements are discussed
in Ref.~\cite{Bru13}. Any approach to these corrections necessarily involve
some model dependence. We have deconvoluted the present measurements by
assigning a mean energy to each point (analogous to Eq.~(8) of
Ref.~\cite{Bru13}) and then applying a correction factor to the
measured yield for each point (analogous to Eq.~(6) of Ref.~\cite{Bru13}).
This procedure requires that the underlying neutron spectrum be
known in advance -- for this we use fit~16. In practice, the fit used
makes very little difference, as long as it gives a reasonable description
of the measured data.
The deconvoluted neutron spectrum data is shown in Fig.~\ref{fig:decon},
where the data from the two detectors are combined and re-binned in energy.

\subsection{$\alpha$-particle spectrum}

Some information about the $\alpha$-particle spectrum from the low-energy
$\text{T}+\text{T}$ reaction is available from a 1985 conference paper by
Jarmie and Brown~\cite{Jar85}, where a measured spectrum and a background
spectrum are given for an incident triton energy of 115~keV and
a laboratory angle of $45^\circ$.
We have extracted the $\alpha$-particle spectrum from their Fig.~8 as follows.
The spectrum was first corrected for the background shown along with
the spectrum in their figure.
Next the spectrum was energy calibrated, using the peak from
the ${\rm T}(d,\alpha)$ reaction to fix the calibration at the high-energy
end of the spectrum. The calibration assumed the channel number in
the Si detector was linear with $\alpha$-particle energy with zero offset,
with the energy loss in the 30-$\mu$g/cm${}^2$ CH$_2$ foil in front of the
detector~\cite{Jar84} being taken into account.
Finally, the spectrum was converted to the
c.m. system assuming the spectrum is isotropic in the c.m. system.
It should be noted that this spectrum should be most reliable for
the higher energies, where the background is small and the energy
calibration is well established.
We finally note that this spectrum was measured for $E_{c.m.}=57.5$~keV,
with the beam-energy loss correction being less than 0.1~keV.

\begin{figure}[p]
\includegraphics[width=0.6\textwidth]{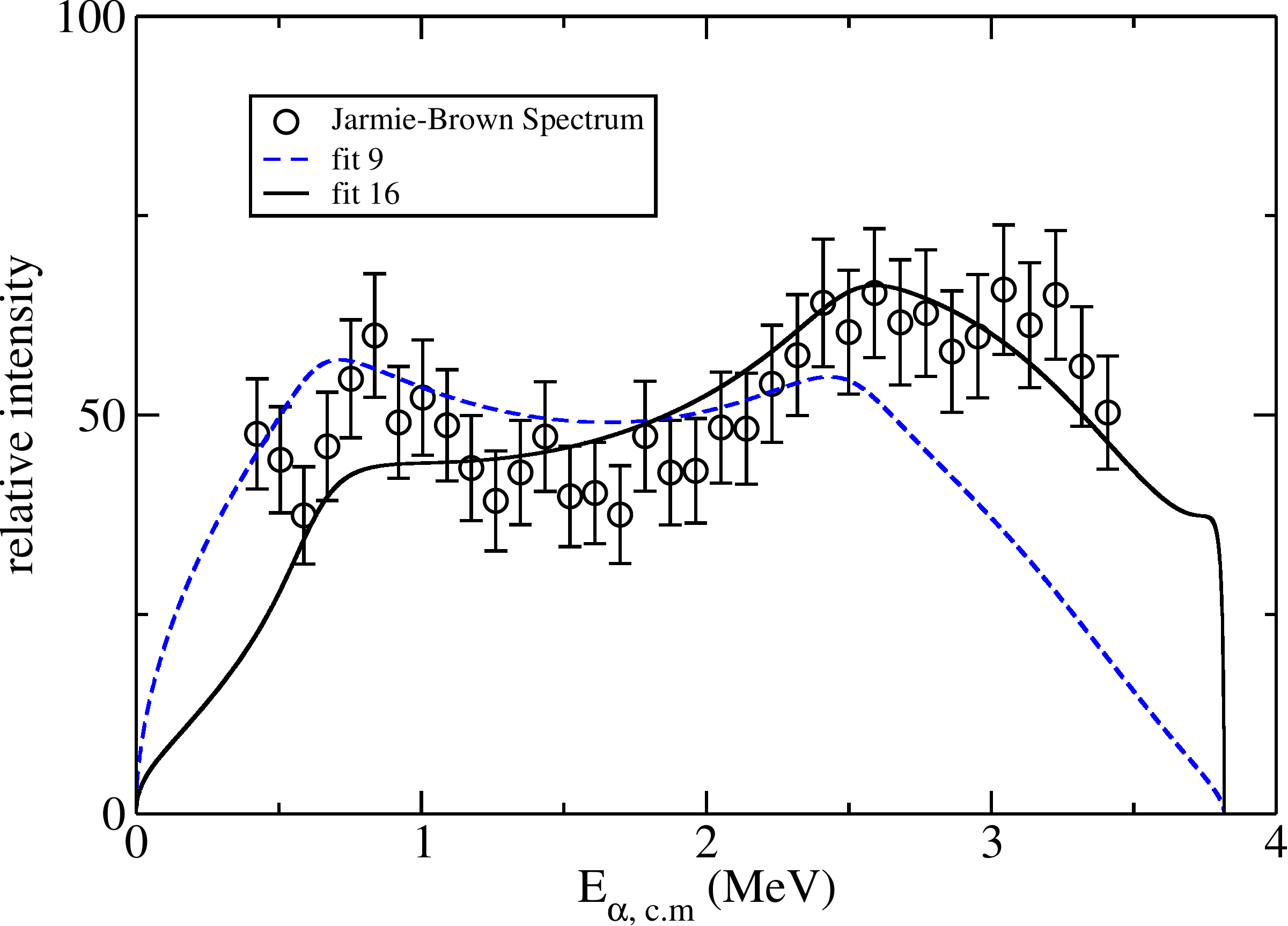}
\caption{\label{fig:jarmie-rebin}The $\alpha$-particle spectrum extracted from
Ref.~\cite{Jar85} (points) and the predictions from fits 9 and~16 (curves).}
\end{figure}

\begin{figure}[p]
\includegraphics[width=0.6\textwidth]{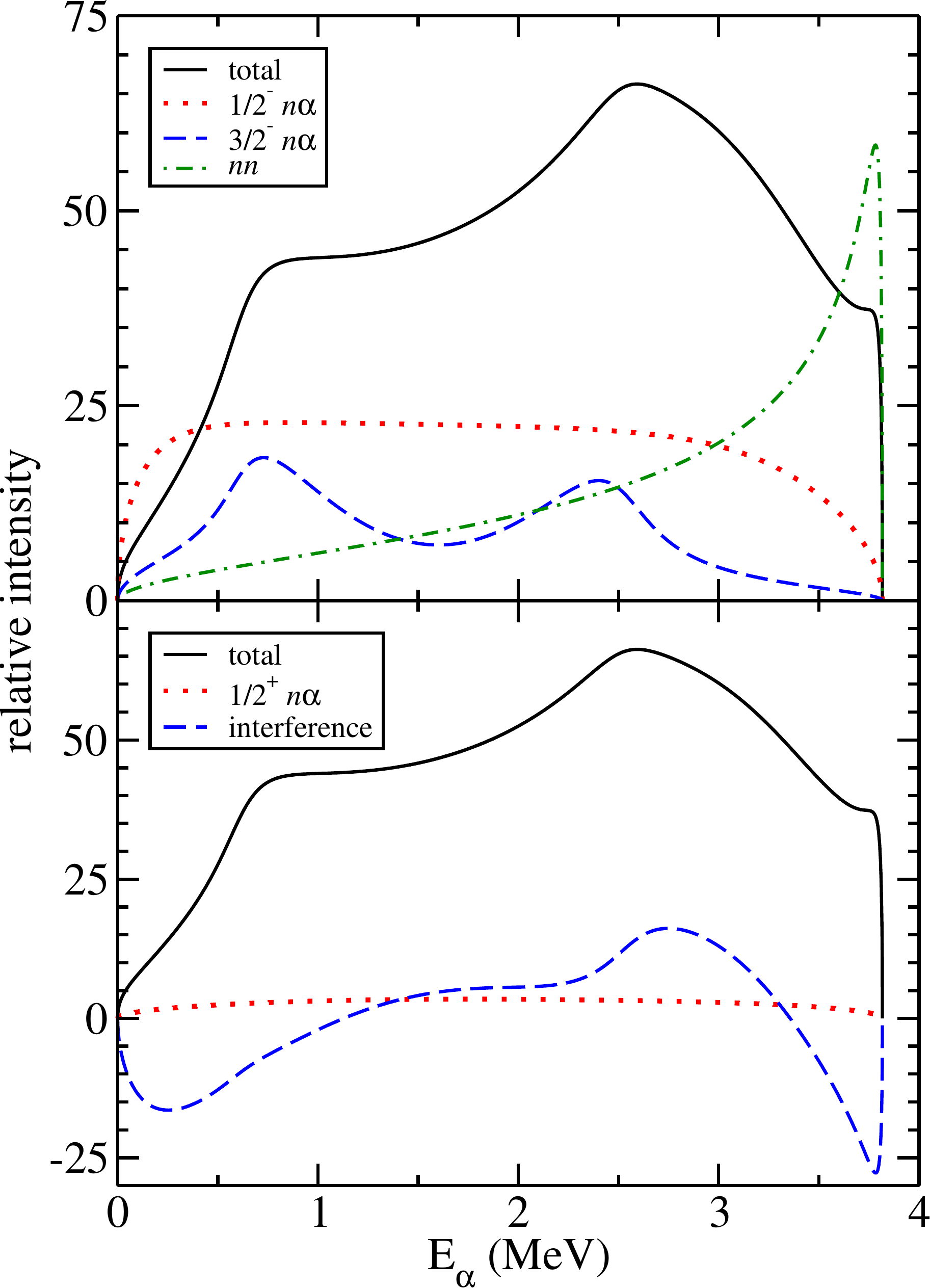}
\caption{\label{fig:adecomp}The decomposition of the $\alpha$-particle
spectrum from the fit~16 prediction into its various
channel components and the net interference contribution.}
\end{figure}

The resulting $\alpha$-particle spectrum, rebinned such that each point
represents 5 channels in the raw spectrum, is shown in
Fig.~\ref{fig:jarmie-rebin}. Also shown are the predictions from $R$-matrix
fits 9 and~16, where the normalizations of the fits has been adjusted to
optimize the agreement with data. It is seen that fit~16 supplies a
much better description of the spectrum than fit~9 ($\chi^2=46$ versus 140
for the 35 data points).
The decomposition of the fit~16 $\alpha$-particle spectrum into its various
channel components and the net interference contribution is shown in
Fig.~\ref{fig:adecomp}.
It is seen that the inclusion of the di-neutron channel, which supplies
spectral strength near the maximum energy, is crucial for reproducing
the spectrum.

\subsection{Dalitz plots}

A useful tool for visualizing the the particle energy correlations in a
three-body final state is the Dalitz plot.
As already noted, the particle distribution given
by Eq.~(\ref{eq:particle_dist}) is a function of two variables.
Taking these to be any pair of particle energies $E_i$ and $E_j$, we can write
\begin{equation}
\frac{d^2N}{dE_idE_j} = \frac{M}{2} \left[ \sum_{c,c'}
  g^{(1)}_{cc'}+g^{(2)}_{cc'}+g^{(12)}_{cc'}\right] ,
\end{equation}
where the kinematically-allowed region in $E_i-E_j$ space is an ellipse.

\begin{figure}[p]
\includegraphics[width=0.6\textwidth]{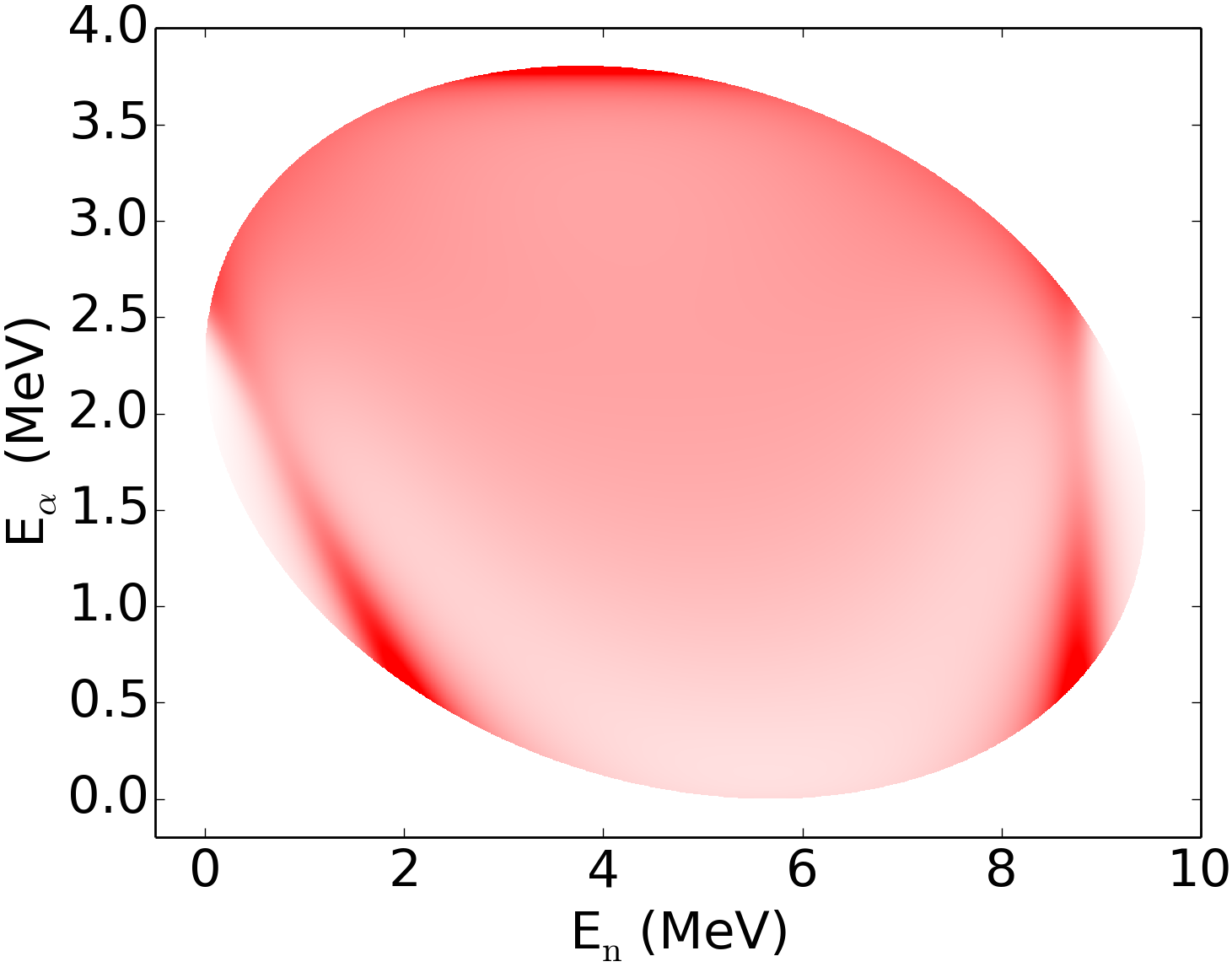}
\caption{\label{fig:tt2n-dalitz}The particle distribution from fit~16
as a function of neutron and $\alpha$-particle energies.}
\end{figure}

The particle distribution resulting from fit~16, plotted as a function
of neutron and $\alpha$-particle energies, is shown in
Fig.~\ref{fig:tt2n-dalitz}. The vertical band at $E_n\approx 8.7$~MeV
and the diagonal band in the lower left part of the ellipse are due
to the ${}^5{\rm He}$ ground state.
The concentration of strength at the top of the
ellipse, where $E_\alpha\approx 3.8$~MeV, is due to the di-neutron.

\begin{figure}[p]
\includegraphics[width=0.6\textwidth]{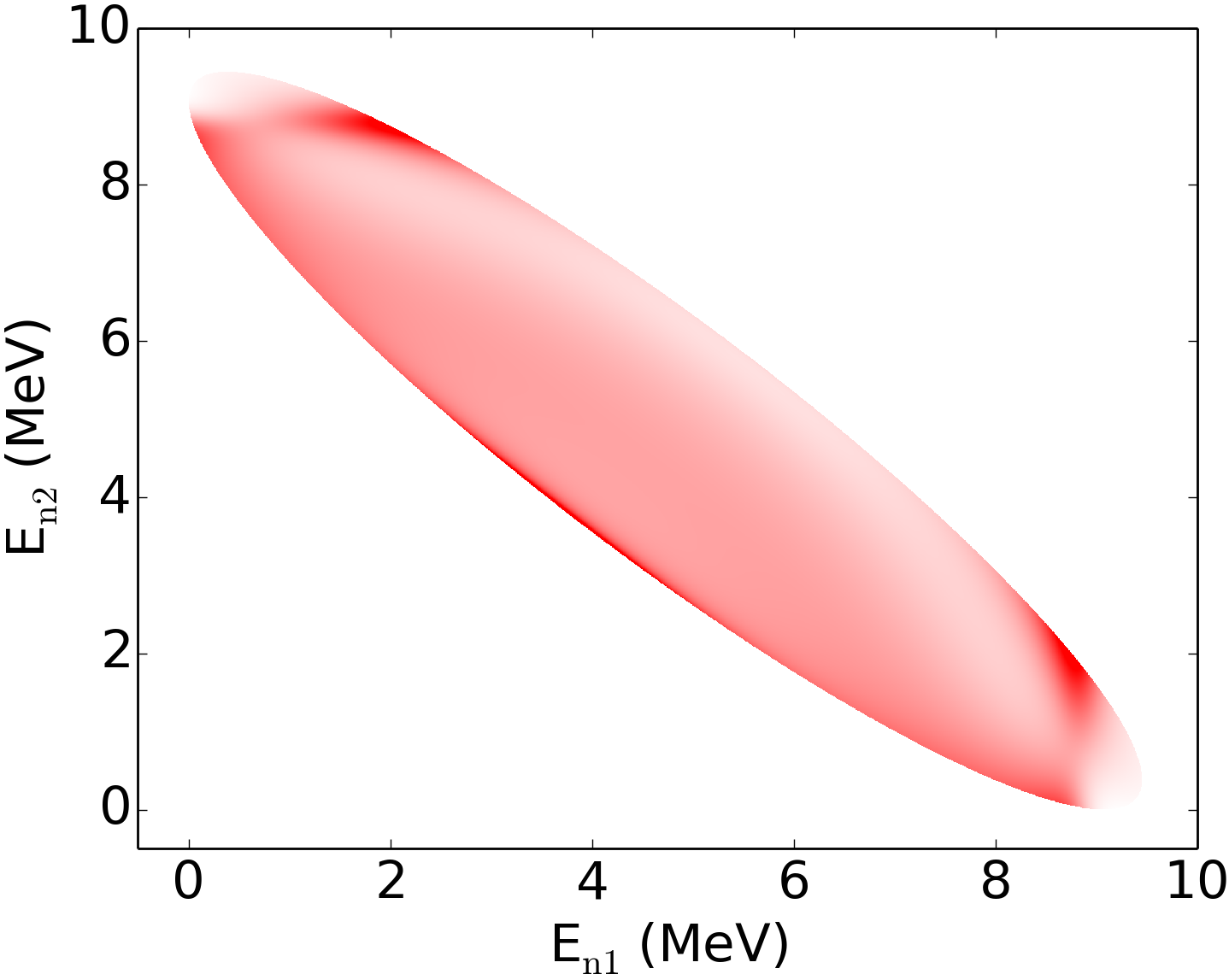}
\caption{\label{fig:tt2n-dalitz-nn}The particle distribution from fit~16
as a function of the two neutron energies.}
\end{figure}

The same particle distribution, plotted as a function of the two
neutron energies, is shown in Fig.~\ref{fig:tt2n-dalitz-nn}.
In this case, the horizontal and vertical bands at $E_n\approx 8.7$~MeV
are due to the ${}^5{\rm He}$ ground state and
the di-neutron strength appears at $E_{n1}\approx E_{n2}\approx 3.8$~MeV.

\section{Particle spectra from ${}^3{\rm He}+{}^3{\rm He}$}
\label{sec:3he_3he}

It is straightforward to adopt this approach for describing the
proton and $\alpha$-particle spectra from  ${}^3{\rm He}+{}^3{\rm He}$,
which is the mirror reaction to $\text{T}+\text{T}\rightarrow 2n+\alpha$.
For the $p\alpha$ channels, we utilize again the final $R$-matrix parameters
given by Ref.~\cite{Sta72}, which are defined using the same channel radius
and boundary condition conventions as their $n\alpha$ parameters used above.
For the $pp$ channel, adopting a channel radius of 2.0~fm, $E_c=4.865$~MeV,
and $\gamma^2_c=34.61$~MeV reproduces the scattering length and effective
range of the Argonne~V18 potential~\cite{Wir95} which are $-7.8064$~fm
and 2.788~fm, respectively.

\begin{figure}[p]
\includegraphics[width=0.6\textwidth]{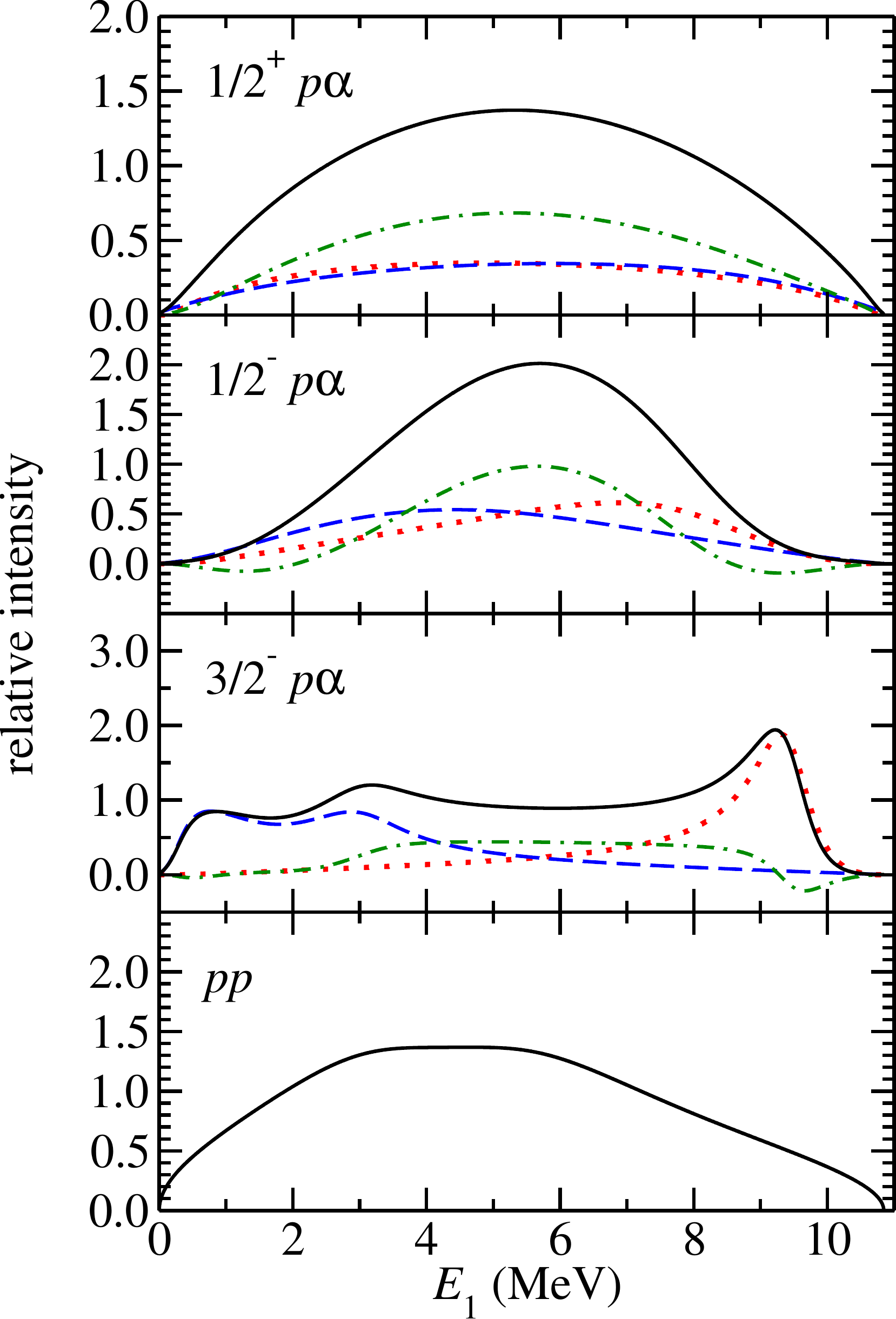}
\caption{\label{fig:pp_dist_chan}Proton energy distributions from
${}^3{\rm He}+{}^3{\rm He}$ for each channel considered separately.
The primary, secondary, exchange, and total are given by the dotted, dashed,
dot-dashed, and solid curves, respectively.
Only the total is shown for the $pp$ case.}
\end{figure}

\begin{figure}[p]
\includegraphics[width=0.6\textwidth]{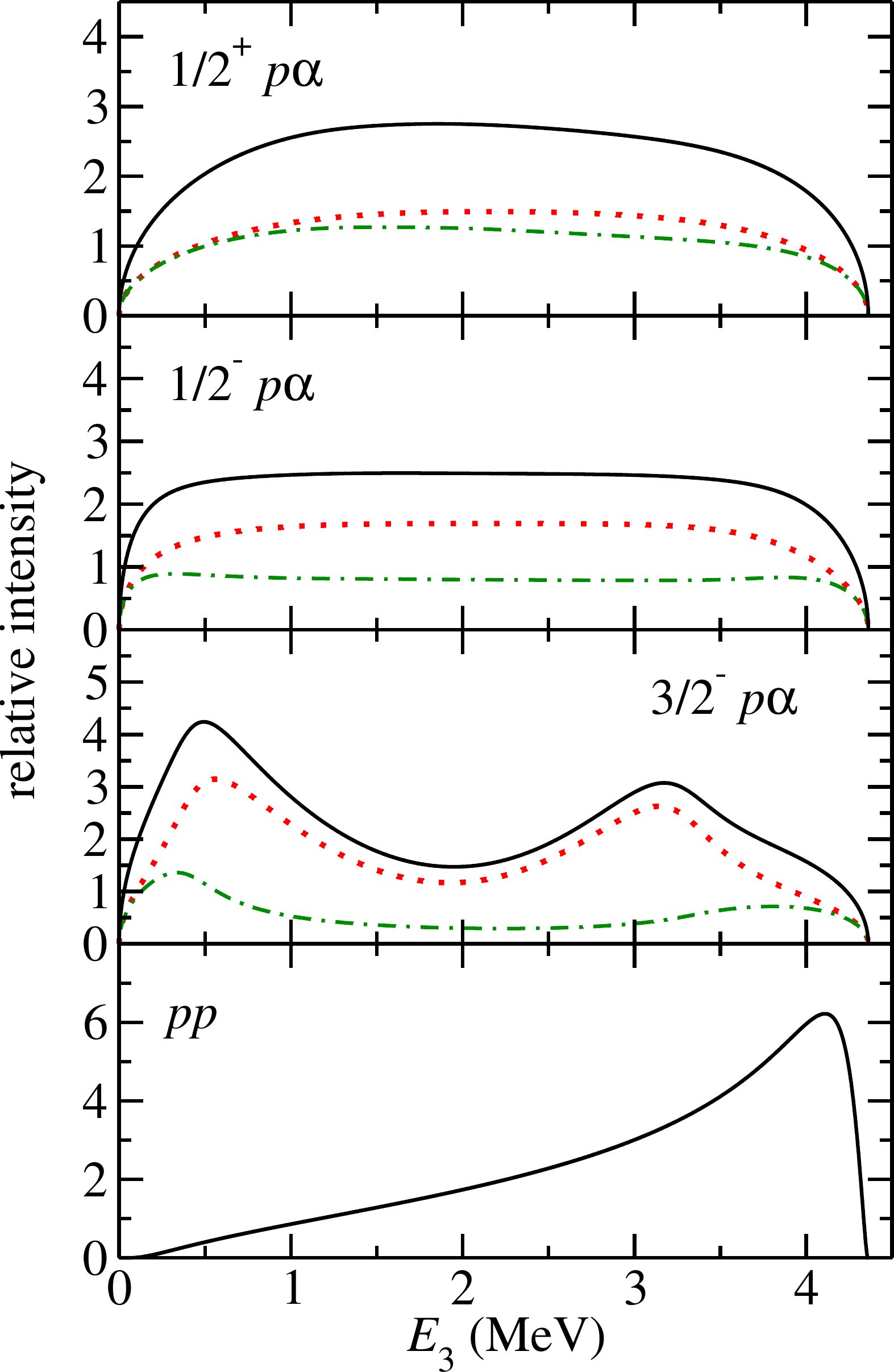}
\caption{\label{fig:a_dist_chan}Alpha-particle energy distributions from
${}^3{\rm He}+{}^3{\rm He}$ for each channel considered separately.
The primary plus secondary, exchange, and total are given by the dotted,
dot-dashed, and solid curves, respectively.
Only the total is shown for the $pp$ case.}
\end{figure}

The calculated results for considering each channel in isolation are
shown in Figs.~\ref{fig:pp_dist_chan} and~\ref{fig:a_dist_chan}.
We have assumed $E_{c.m.}=165$~keV and the same normalization convention was
used as for the $\text{T}+\text{T}$ calculations shown in
Figs.~\ref{fig:ndist_chan} and~\ref{fig:adist_chan}.
The results are very similar to those found for $\text{T}+\text{T}$.
The main difference is that the ${}^5{\rm Li}$ ground-state peak in
$3/2^-$ $p\alpha$ channel is broader than the ${}^5{\rm He}$ peak in the
corresponding $n\alpha$ channel, which simply reflects the different
intrinsic widths of the states. In addition, the $\alpha$-particle
energy distribution is somewhat broader for the $pp$ channel compared
to the corresponding $nn$ channel.

\begin{figure}[p]
\includegraphics[width=0.6\textwidth]{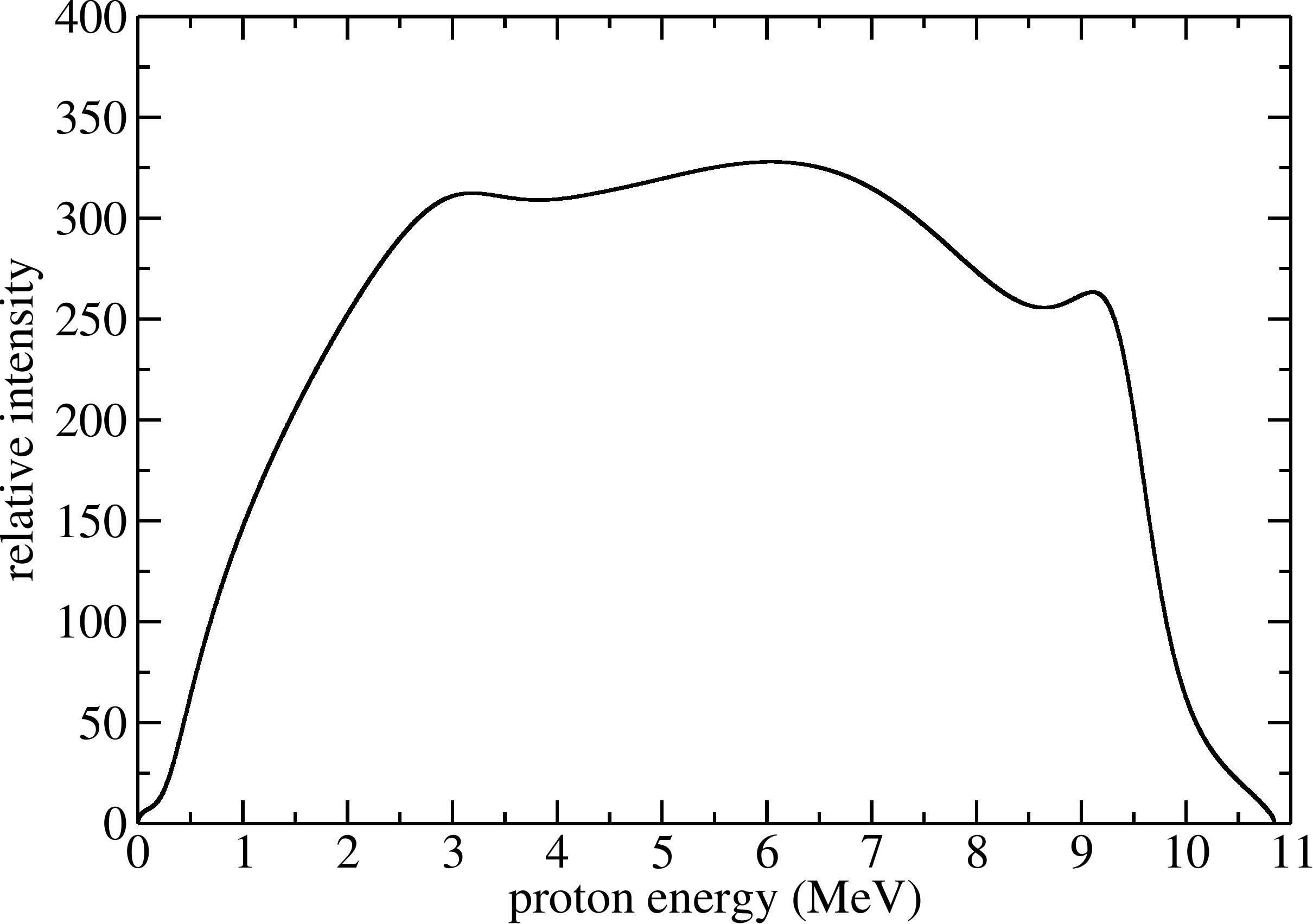}
\caption{\label{fig:3he-3he_sayre}The predicted ${}^3{\rm He}+{}^3{\rm He}$
proton spectrum for $E_{c.m.}=165$~keV.}
\end{figure}

\begin{figure}[p]
\includegraphics[width=0.6\textwidth]{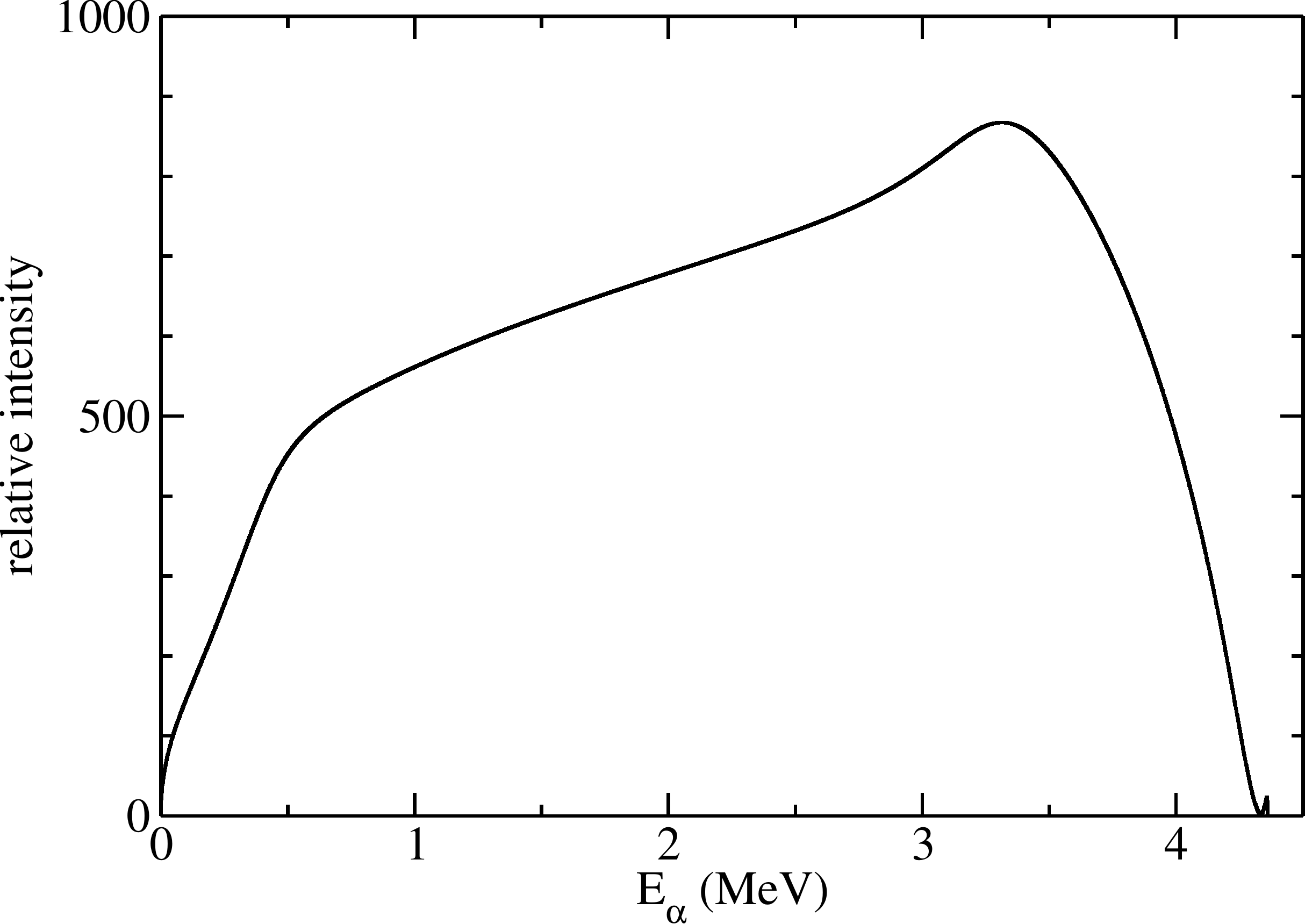}
\caption{\label{fig:3he-3he_alpha}The predicted ${}^3{\rm He}+{}^3{\rm He}$
$\alpha$-particle spectrum for $E_{c.m.}=165$~keV.}
\end{figure}

In Fig.~\ref{fig:3he-3he_sayre}, we show a prediction for the
proton spectrum, where we have assumed the feeding factors from
fit~16 to $\text{T}+\text{T}$ neutron spectrum.
The corresponding prediction for the $\alpha$-particle spectrum
is shown in Fig.~\ref{fig:3he-3he_alpha}.
The main differences from the $\text{T}+\text{T}$ case are
(1) the ${}^5{\rm Li}$ ground state is less prominent than the
${}^5{\rm He}$ ground state peak in the proton spectrum,
due to the former being broader, and
(2) there is less strength in the $\alpha$-particle spectrum near
the endpoint (di-proton region), due to the Coulomb barrier between
the two protons suppressing the amplitude as their relative energy
approaches zero.

It should be noted that we have ignored certain complications introduced by
long-ranged Coulomb force to the three-body final state.
In particular, our factorized form of the amplitude does {\em not}
correspond to an asymptotic solution to the Schr{\" o}dinger equation
for three charged particles~\cite{Alt93,Muk96}.
One can see that our amplitude does not include the
effect of the Coulomb barrier as the relative energy goes to zero
for {\em all} particle pairs.
For some energy spectra and channels, this deficiency is exposed at the
highest energies in the particle spectra, as the endpoint corresponds to
the case where the other two particles recoil in the opposite direction
with zero relative energy.

\begin{figure}[p]
\includegraphics[width=0.6\textwidth]{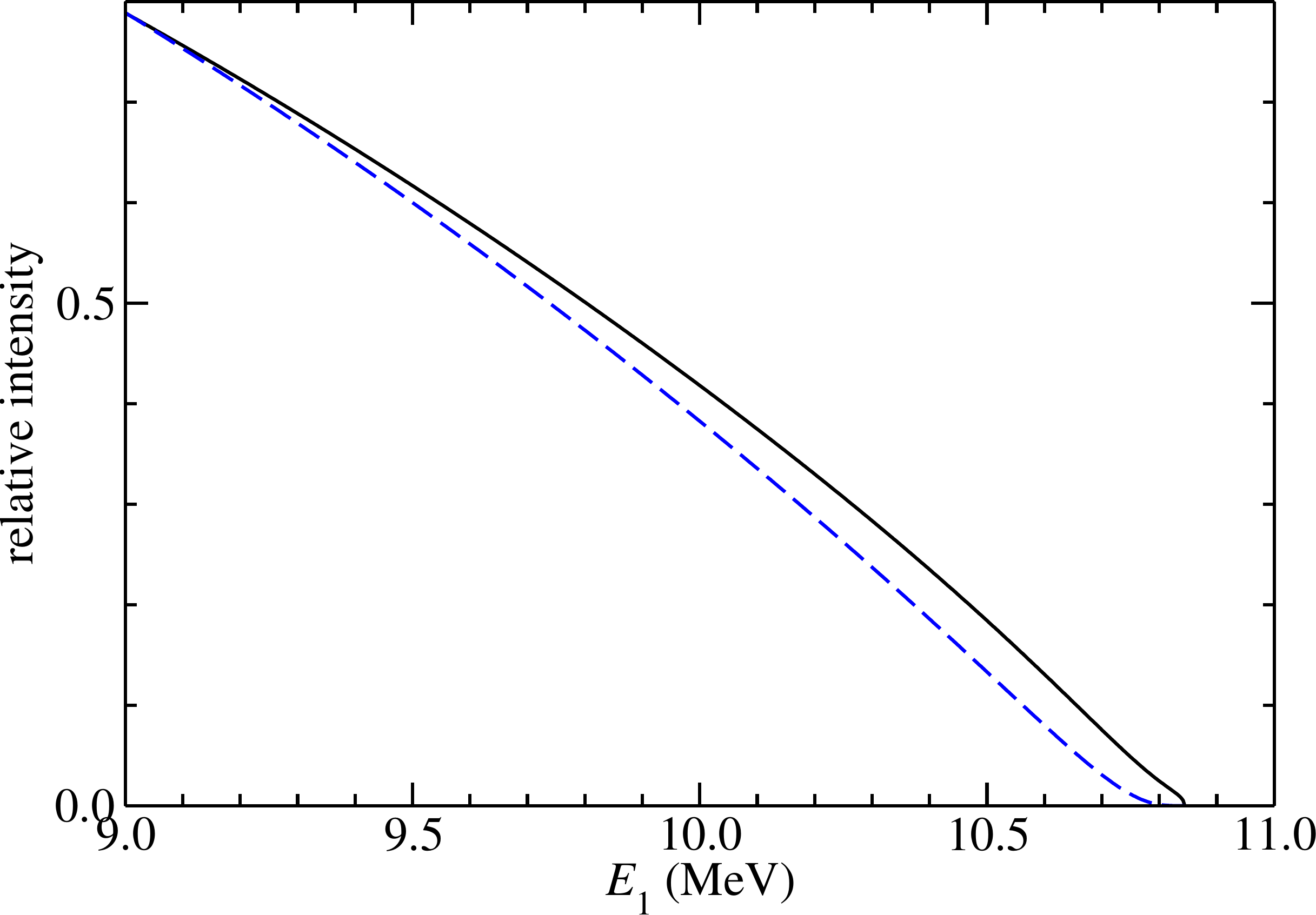}
\caption{\label{fig:ppa-endpoint}Proton energy distributions from
${}^3{\rm He}+{}^3{\rm He}$ for the $1/2^+$ $p\alpha$ channel near the
endpoint.
The solid curve is the same as shown in Fig.~\ref{fig:pp_dist_chan} and
the dashed curve shows the effect of including the ad-hoc Coulomb
correction discussed in the text.}
\end{figure}

An {\em ad hoc} modification to our amplitudes can be made which restores
physically-reasonable behavior near the endpoints.
Such an approach may be necessary for describing experimental data in
these regions.
A simple procedure is to multiply the $p\alpha$ amplitude given by
Eq.~(\ref{eq:u_n_alpha})  by $C_{12}C_{13}$ and
the $pp$ amplitude given by Eq.~(\ref{eq:u_n_n}) by $C_{13}C_{23}$, where
\begin{equation}
C_{ij}=\left[\frac{P_0(k_{ij}a_{ij},\eta_{ij})}{P_0(k_{ij}a_{ij},0)}
  \right]^{1/2},
\end{equation}
and $k_{ij}$, $a_{ij}$, and $\eta_{ij}$ are the wavenumber, channel
radius, and Coulomb parameter for particle pair $ij$. We have assumed 
$l=0$ for the penetration factor and used the same radii for the
$p\alpha$ and $pp$ channels as discussed above.
This modification introduces some additional angular dependence to
the matrix element that prevents some of the simplifications based on
integrating over Legendre polynomials discussed in
Subsec.~\ref{subsec:calc_spec} from being applicable.
Otherwise, the computations are unchanged.
We find that the shapes of the calculated spectra are little changed except
near the endpoint while the normalization (area) of the spectra are
reduced by 12-17\%, depending up on the particular channel.
The results with and without this modification are shown in
Fig.~\ref{fig:ppa-endpoint} for the $1/2^+$ $p\alpha$ channel near the
endpoint, where the normalization of the
modified spectrum was adjusted to match the area of the original spectrum.

\section{Heavy nuclei}
\label{sec:heavy}

It is instructive to consider how this formalism behaves in heavier
nuclei. In the limit that $m_3 \gg m_1,m_2$, the kinematic relations
simplify considerably. In particular, we have $E_1+E_2=E_{tot}$, $p_{23}=p_2$,
$p_{13}=p_1$, and $\cos\gamma_{23}=\cos\gamma_{13}=\cos\delta_{12}$.
The is also substantial simplification of the angular functions:
\begin{equation}
W^{(1)}_{lJl'J'} = W^{(2)}_{lJl'J'} = -W^{(12)}_{lJl'J'}.
\end{equation}
If only a single $n\alpha$ channel is present, the neutron energy spectrum
is given by
\begin{equation}
\frac{dN}{dE_1}=p_1p_2\left|u_c(E_1)+u_c(E_{tot}-E_1)\right|^2,
\end{equation}
which is symmetric around the center of the spectrum $(E_1=E_{tot}/2)$.
Note also that the interference due to antisymmetrization is
maximally constructive at the center of the spectrum.
This result explains the general tendency observed in
Fig.~\ref{fig:ndist_chan} for the antisymmetrization interference
contribution to the neutron spectrum to be constructive
in the $\text{T}+\text{T}$ case.
This formula was determined using the following special result for
the Racah coefficient~\cite{Dev57} when $k=0$
\begin{equation}
W(0lJ'\frac{1}{2};l'J)=\frac{\delta_{JJ'}\delta_{ll'}}
  {\sqrt{(2J+1)(2l+1)}}.
\end{equation}
This result also implies that the interference between channels with
distinct $l$ and $J$ values vanishes in this limit.

The interference between channels with distinct $l$ and $J$ values
can thus be interpreted to arise via the recoil of the
intermediate state. This recoil is substantial in processes
involving light nuclei such as $\text{T}+\text{T}\rightarrow 2n+\alpha$.
In heavier nuclei, the interference between channels is
much reduced and is found to scale $\propto 1/m_3$.

\section{Conclusions}

A phenomenological $R$-matrix model for the three-body final state of the
$\text{T}+\text{T}\rightarrow 2n+\alpha$ reaction has been presented.
This approach includes a detailed treatment of the $n\alpha$ and
$nn$ interactions in the final state, angular momentum conservation,
antisymmetrization, and the interference between different channels.
This model is able to supply an excellent fit to the $\text{T}+\text{T}$
neutron spectrum for $E_{c.m.}=16$~keV recently measured at the
NIF~\cite{Say13}.
The most prominent feature in the spectrum is a peak at
$E_n\approx 8.7$~MeV, which arises for the $3/2^-$ ${}^5{\rm He}$ ground state.
The strength in the spectrum at lower neutron energies arises from
the $1/2^-$ first excited state of ${}^5{\rm He}$,
$1/2^+$ $n\alpha$ emission, and the $nn$ (di-neutron) emission channel.
The best fit to the spectrum includes significant strength in the di-neutron
channel, but it should be noted that the distribution of strengths in these
additional channels is not well constrained by the data
(see Table~\ref{tab:fit_results}).
This best fit provides a prediction for the $\alpha$-particle spectrum,
which is in reasonable agreement with an experimental spectrum that is
available in the literature~\cite{Jar85}.
The agreement of the prediction with the data near the endpoint of the
$\alpha$-particle spectrum provides support for the significant di-neutron
channel strength present in the best fit.

Several issues could be clarified by improved experimental data.
It would be very useful to extend the $\text{T}+\text{T}$ neutron spectrum
measurements to lower neutron energies, in order to better constrain the
fits and to possibly observe the double-humped structured predicted below
2~MeV neutron energy that is associated with the ${}^5{\rm He}$ ground state
(see the $3/2^-$ $n\alpha$ panel in Fig.~\ref{fig:ndist_chan}).
A fully documented measurement of the $\alpha$-particle spectrum from
$\text{T}+\text{T}$ would also be valuable, particularly if the spectrum
could be measured up to the endpoint, where the di-neutron contribution
is maximal. It would also be interesting to study the dependence of
the spectrum on the energy in the entrance channel, as there is some
indication the ${}^5{\rm He}$ ground state peak is more prominent
at higher entrance channel energies~\cite{Won65}.

It is also interesting to consider the reactions
${}^3\text{He}+{}^3\text{He}\rightarrow 2p+\alpha$ and
$\text{T}+{}^3\text{He}\rightarrow n+p+\alpha$, which are
related by mirror or isospin symmetry to
$\text{T}+\text{T}\rightarrow 2n+\alpha$.
A prediction for the proton and $\alpha$ spectra resulting from
${}^3\text{He}+{}^3\text{He}$ has been given above in
Sec.~\ref{sec:3he_3he}.
Measurements of proton spectra from ${}^3\text{He}+{}^3\text{He}$ and
$\text{T}+{}^3\text{He}$ are currently being pursued with the inertial
confinement fusion technique using the OMEGA facility at the Laboratory for
Laser Energetics of the University of Rochester~\cite{Zyl14}.

On the theoretical side, it would be interesting and useful to extend
the formalism presented here to include the energy dependence in the
initial state. We expect that the methods presented here can be applied
to additional reactions or spectra with three-body final states.
For example, our approach could be applied to the decay
${}^{16}{\rm Be}\rightarrow 2n+{}^{14}{\rm Be}$,
where evidence for the di-neutron has been reported~\cite{Spy12}.
Another area where these methods could be used is the calculation
of coherent interference effects between different decay pathways
to three-body final states, which has been noted as an important
issue for understanding the total widths of states which decay
by the emission of three particles~\cite{Bar03,Bar08}.
It must also be acknowledged that the phenomenological
$R$-matrix approach presented here includes many approximations.
In the future, it is thus hoped that ab-initio techniques based on
nucleon-nucleon interactions may be applied to these reactions.

\begin{acknowledgments}

We thank Johan Frenje, Maria Gatu-Johnson, Dennis McNabb, Dieter Schneider,
Ian Thompson, and Alex Zylstra for useful discussions.
The work was supported in part by the U.S. Department of Energy,
under Grant Nos. DE-FG02-88ER40387, DE-NA0001837, and DE-AC52-06NA25396,
and by Lawrence Livermore National Laboratory.

\end{acknowledgments}

\bibliography{tt2n}

\begin{thebibliography}{26}%
\makeatletter
\providecommand \@ifxundefined [1]{%
 \@ifx{#1\undefined}
}%
\providecommand \@ifnum [1]{%
 \ifnum #1\expandafter \@firstoftwo
 \else \expandafter \@secondoftwo
 \fi
}%
\providecommand \@ifx [1]{%
 \ifx #1\expandafter \@firstoftwo
 \else \expandafter \@secondoftwo
 \fi
}%
\providecommand \natexlab [1]{#1}%
\providecommand \enquote  [1]{``#1''}%
\providecommand \bibnamefont  [1]{#1}%
\providecommand \bibfnamefont [1]{#1}%
\providecommand \citenamefont [1]{#1}%
\providecommand \href@noop [0]{\@secondoftwo}%
\providecommand \href [0]{\begingroup \@sanitize@url \@href}%
\providecommand \@href[1]{\@@startlink{#1}\@@href}%
\providecommand \@@href[1]{\endgroup#1\@@endlink}%
\providecommand \@sanitize@url [0]{\catcode `\\12\catcode `\$12\catcode
  `\&12\catcode `\#12\catcode `\^12\catcode `\_12\catcode `\%12\relax}%
\providecommand \@@startlink[1]{}%
\providecommand \@@endlink[0]{}%
\providecommand \url  [0]{\begingroup\@sanitize@url \@url }%
\providecommand \@url [1]{\endgroup\@href {#1}{\urlprefix }}%
\providecommand \urlprefix  [0]{URL }%
\providecommand \Eprint [0]{\href }%
\providecommand \doibase [0]{http://dx.doi.org/}%
\providecommand \selectlanguage [0]{\@gobble}%
\providecommand \bibinfo  [0]{\@secondoftwo}%
\providecommand \bibfield  [0]{\@secondoftwo}%
\providecommand \translation [1]{[#1]}%
\providecommand \BibitemOpen [0]{}%
\providecommand \bibitemStop [0]{}%
\providecommand \bibitemNoStop [0]{.\EOS\space}%
\providecommand \EOS [0]{\spacefactor3000\relax}%
\providecommand \BibitemShut  [1]{\csname bibitem#1\endcsname}%
\let\auto@bib@innerbib\@empty
\bibitem [{\citenamefont {Sayre}\ \emph {et~al.}(2013)\citenamefont {Sayre},
  \citenamefont {Brune}, \citenamefont {Caggiano}, \citenamefont {Glebov},
  \citenamefont {Hatarik}, \citenamefont {Bacher}, \citenamefont {Bleuel},
  \citenamefont {Casey}, \citenamefont {Cerjan}, \citenamefont {Eckart},
  \citenamefont {Fortner}, \citenamefont {Frenje}, \citenamefont {Friedrich},
  \citenamefont {Gatu-Johnson}, \citenamefont {Grim}, \citenamefont {Hagmann},
  \citenamefont {Knauer}, \citenamefont {Kline}, \citenamefont {McNabb},
  \citenamefont {McNaney}, \citenamefont {Mintz}, \citenamefont {Moran},
  \citenamefont {Nikroo}, \citenamefont {Phillips}, \citenamefont {Pino},
  \citenamefont {Remington}, \citenamefont {Rowley}, \citenamefont {Schneider},
  \citenamefont {Smalyuk}, \citenamefont {Stoeffl}, \citenamefont {Tipton},
  \citenamefont {Weber},\ and\ \citenamefont {Yeamans}}]{Say13}%
  \BibitemOpen
  \bibfield  {author} {\bibinfo {author} {\bibfnamefont {D.~B.}\ \bibnamefont
  {Sayre}}, \bibinfo {author} {\bibfnamefont {C.~R.}\ \bibnamefont {Brune}},
  \bibinfo {author} {\bibfnamefont {J.~A.}\ \bibnamefont {Caggiano}}, \bibinfo
  {author} {\bibfnamefont {V.~Y.}\ \bibnamefont {Glebov}}, \bibinfo {author}
  {\bibfnamefont {R.}~\bibnamefont {Hatarik}}, \bibinfo {author} {\bibfnamefont
  {A.~D.}\ \bibnamefont {Bacher}}, \bibinfo {author} {\bibfnamefont {D.~L.}\
  \bibnamefont {Bleuel}}, \bibinfo {author} {\bibfnamefont {D.~T.}\
  \bibnamefont {Casey}}, \bibinfo {author} {\bibfnamefont {C.~J.}\ \bibnamefont
  {Cerjan}}, \bibinfo {author} {\bibfnamefont {M.~J.}\ \bibnamefont {Eckart}},
  \bibinfo {author} {\bibfnamefont {R.~J.}\ \bibnamefont {Fortner}}, \bibinfo
  {author} {\bibfnamefont {J.~A.}\ \bibnamefont {Frenje}}, \bibinfo {author}
  {\bibfnamefont {S.}~\bibnamefont {Friedrich}}, \bibinfo {author}
  {\bibfnamefont {M.}~\bibnamefont {Gatu-Johnson}}, \bibinfo {author}
  {\bibfnamefont {G.~P.}\ \bibnamefont {Grim}}, \bibinfo {author}
  {\bibfnamefont {C.}~\bibnamefont {Hagmann}}, \bibinfo {author} {\bibfnamefont
  {J.~P.}\ \bibnamefont {Knauer}}, \bibinfo {author} {\bibfnamefont {J.~L.}\
  \bibnamefont {Kline}}, \bibinfo {author} {\bibfnamefont {D.~P.}\ \bibnamefont
  {McNabb}}, \bibinfo {author} {\bibfnamefont {J.~M.}\ \bibnamefont {McNaney}},
  \bibinfo {author} {\bibfnamefont {J.~M.}\ \bibnamefont {Mintz}}, \bibinfo
  {author} {\bibfnamefont {M.~J.}\ \bibnamefont {Moran}}, \bibinfo {author}
  {\bibfnamefont {A.}~\bibnamefont {Nikroo}}, \bibinfo {author} {\bibfnamefont
  {T.}~\bibnamefont {Phillips}}, \bibinfo {author} {\bibfnamefont {J.~E.}\
  \bibnamefont {Pino}}, \bibinfo {author} {\bibfnamefont {B.~A.}\ \bibnamefont
  {Remington}}, \bibinfo {author} {\bibfnamefont {D.~P.}\ \bibnamefont
  {Rowley}}, \bibinfo {author} {\bibfnamefont {D.~H.}\ \bibnamefont
  {Schneider}}, \bibinfo {author} {\bibfnamefont {V.~A.}\ \bibnamefont
  {Smalyuk}}, \bibinfo {author} {\bibfnamefont {W.}~\bibnamefont {Stoeffl}},
  \bibinfo {author} {\bibfnamefont {R.~E.}\ \bibnamefont {Tipton}}, \bibinfo
  {author} {\bibfnamefont {S.~V.}\ \bibnamefont {Weber}}, \ and\ \bibinfo
  {author} {\bibfnamefont {C.~B.}\ \bibnamefont {Yeamans}},\ }\href {\doibase
  10.1103/PhysRevLett.111.052501} {\bibfield  {journal} {\bibinfo  {journal}
  {Phys. Rev. Lett.}\ }\textbf {\bibinfo {volume} {111}},\ \bibinfo {pages}
  {052501} (\bibinfo {year} {2013})}\BibitemShut {NoStop}%
\bibitem [{\citenamefont {Ohlsen}(1965)}]{Ohl65}%
  \BibitemOpen
  \bibfield  {author} {\bibinfo {author} {\bibfnamefont {G.~G.}\ \bibnamefont
  {Ohlsen}},\ }\href {\doibase 10.1016/0029-554X(65)90368-X} {\bibfield
  {journal} {\bibinfo  {journal} {Nuclear Instruments and Methods}\ }\textbf
  {\bibinfo {volume} {37}},\ \bibinfo {pages} {240 } (\bibinfo {year}
  {1965})}\BibitemShut {NoStop}%
\bibitem [{\citenamefont {Barker}(1988)}]{Bar88}%
  \BibitemOpen
  \bibfield  {author} {\bibinfo {author} {\bibfnamefont {F.~C.}\ \bibnamefont
  {Barker}},\ }\href@noop {} {\bibfield  {journal} {\bibinfo  {journal}
  {Australian Journal of Physics}\ }\textbf {\bibinfo {volume} {41}},\ \bibinfo
  {pages} {743} (\bibinfo {year} {1988})}\BibitemShut {NoStop}%
\bibitem [{\citenamefont {Lacina}\ \emph {et~al.}(1965)\citenamefont {Lacina},
  \citenamefont {Ingley},\ and\ \citenamefont {Dorn}}]{Lac65}%
  \BibitemOpen
  \bibfield  {author} {\bibinfo {author} {\bibfnamefont {B.}~\bibnamefont
  {Lacina}}, \bibinfo {author} {\bibfnamefont {J.}~\bibnamefont {Ingley}}, \
  and\ \bibinfo {author} {\bibfnamefont {D.~W.}\ \bibnamefont {Dorn}},\
  }\href@noop {} {\emph {\bibinfo {title} {Neutron-neutron interaction in the
  T+T reaction}}},\ \bibinfo {type} {Tech. Rep.}\ \bibinfo {number}
  {UCRL-7769}\ (\bibinfo  {institution} {Lawrence Radiation Laboratory},\
  \bibinfo {year} {1965})\BibitemShut {NoStop}%
\bibitem [{\citenamefont {Balamuth}\ \emph {et~al.}(1974)\citenamefont
  {Balamuth}, \citenamefont {Zurm\"uhle},\ and\ \citenamefont {Tabor}}]{Bal74}%
  \BibitemOpen
  \bibfield  {author} {\bibinfo {author} {\bibfnamefont {D.~P.}\ \bibnamefont
  {Balamuth}}, \bibinfo {author} {\bibfnamefont {R.~W.}\ \bibnamefont
  {Zurm\"uhle}}, \ and\ \bibinfo {author} {\bibfnamefont {S.~L.}\ \bibnamefont
  {Tabor}},\ }\href {\doibase 10.1103/PhysRevC.10.975} {\bibfield  {journal}
  {\bibinfo  {journal} {Phys. Rev. C}\ }\textbf {\bibinfo {volume} {10}},\
  \bibinfo {pages} {975} (\bibinfo {year} {1974})}\BibitemShut {NoStop}%
\bibitem [{\citenamefont {Geesaman}\ \emph {et~al.}(1977)\citenamefont
  {Geesaman}, \citenamefont {McGrath}, \citenamefont {Lesser}, \citenamefont
  {Urone},\ and\ \citenamefont {VerWest}}]{Gee77}%
  \BibitemOpen
  \bibfield  {author} {\bibinfo {author} {\bibfnamefont {D.~F.}\ \bibnamefont
  {Geesaman}}, \bibinfo {author} {\bibfnamefont {R.~L.}\ \bibnamefont
  {McGrath}}, \bibinfo {author} {\bibfnamefont {P.~M.~S.}\ \bibnamefont
  {Lesser}}, \bibinfo {author} {\bibfnamefont {P.~P.}\ \bibnamefont {Urone}}, \
  and\ \bibinfo {author} {\bibfnamefont {B.}~\bibnamefont {VerWest}},\ }\href
  {\doibase 10.1103/PhysRevC.15.1835} {\bibfield  {journal} {\bibinfo
  {journal} {Phys. Rev. C}\ }\textbf {\bibinfo {volume} {15}},\ \bibinfo
  {pages} {1835} (\bibinfo {year} {1977})}\BibitemShut {NoStop}%
\bibitem [{\citenamefont {Fynbo}\ \emph {et~al.}(2003)\citenamefont {Fynbo},
  \citenamefont {Prezado}, \citenamefont {Bergmann}, \citenamefont {Borge},
  \citenamefont {Dendooven}, \citenamefont {Huang}, \citenamefont {Huikari},
  \citenamefont {Jeppesen}, \citenamefont {Jones}, \citenamefont {Jonson},
  \citenamefont {Meister}, \citenamefont {Nyman}, \citenamefont {Riisager},
  \citenamefont {Tengblad}, \citenamefont {Vogelius}, \citenamefont {Wang},
  \citenamefont {Weissman}, \citenamefont {Rolander},\ and\ \citenamefont
  {\"Ayst\"o}}]{Fyn03}%
  \BibitemOpen
  \bibfield  {author} {\bibinfo {author} {\bibfnamefont {H.~O.~U.}\
  \bibnamefont {Fynbo}}, \bibinfo {author} {\bibfnamefont {Y.}~\bibnamefont
  {Prezado}}, \bibinfo {author} {\bibfnamefont {U.~C.}\ \bibnamefont
  {Bergmann}}, \bibinfo {author} {\bibfnamefont {M.~J.~G.}\ \bibnamefont
  {Borge}}, \bibinfo {author} {\bibfnamefont {P.}~\bibnamefont {Dendooven}},
  \bibinfo {author} {\bibfnamefont {W.~X.}\ \bibnamefont {Huang}}, \bibinfo
  {author} {\bibfnamefont {J.}~\bibnamefont {Huikari}}, \bibinfo {author}
  {\bibfnamefont {H.}~\bibnamefont {Jeppesen}}, \bibinfo {author}
  {\bibfnamefont {P.}~\bibnamefont {Jones}}, \bibinfo {author} {\bibfnamefont
  {B.}~\bibnamefont {Jonson}}, \bibinfo {author} {\bibfnamefont
  {M.}~\bibnamefont {Meister}}, \bibinfo {author} {\bibfnamefont
  {G.}~\bibnamefont {Nyman}}, \bibinfo {author} {\bibfnamefont
  {K.}~\bibnamefont {Riisager}}, \bibinfo {author} {\bibfnamefont
  {O.}~\bibnamefont {Tengblad}}, \bibinfo {author} {\bibfnamefont {I.~S.}\
  \bibnamefont {Vogelius}}, \bibinfo {author} {\bibfnamefont {Y.}~\bibnamefont
  {Wang}}, \bibinfo {author} {\bibfnamefont {L.}~\bibnamefont {Weissman}},
  \bibinfo {author} {\bibfnamefont {K.~W.}\ \bibnamefont {Rolander}}, \ and\
  \bibinfo {author} {\bibfnamefont {J.}~\bibnamefont {\"Ayst\"o}},\ }\href
  {\doibase 10.1103/PhysRevLett.91.082502} {\bibfield  {journal} {\bibinfo
  {journal} {Phys. Rev. Lett.}\ }\textbf {\bibinfo {volume} {91}},\ \bibinfo
  {pages} {082502} (\bibinfo {year} {2003})}\BibitemShut {NoStop}%
\bibitem [{\citenamefont {Fynbo}\ \emph {et~al.}(2009)\citenamefont {Fynbo},
  \citenamefont {\'Alvarez-Rodr\'{\i}guez}, \citenamefont {Jensen},
  \citenamefont {Kirsebom}, \citenamefont {Fedorov},\ and\ \citenamefont
  {Garrido}}]{Fyn09}%
  \BibitemOpen
  \bibfield  {author} {\bibinfo {author} {\bibfnamefont {H.~O.~U.}\
  \bibnamefont {Fynbo}}, \bibinfo {author} {\bibfnamefont {R.}~\bibnamefont
  {\'Alvarez-Rodr\'{\i}guez}}, \bibinfo {author} {\bibfnamefont {A.~S.}\
  \bibnamefont {Jensen}}, \bibinfo {author} {\bibfnamefont {O.~S.}\
  \bibnamefont {Kirsebom}}, \bibinfo {author} {\bibfnamefont {D.~V.}\
  \bibnamefont {Fedorov}}, \ and\ \bibinfo {author} {\bibfnamefont
  {E.}~\bibnamefont {Garrido}},\ }\href {\doibase 10.1103/PhysRevC.79.054009}
  {\bibfield  {journal} {\bibinfo  {journal} {Phys. Rev. C}\ }\textbf {\bibinfo
  {volume} {79}},\ \bibinfo {pages} {054009} (\bibinfo {year}
  {2009})}\BibitemShut {NoStop}%
\bibitem [{\citenamefont {Stammbach}\ and\ \citenamefont
  {Walter}(1972)}]{Sta72}%
  \BibitemOpen
  \bibfield  {author} {\bibinfo {author} {\bibfnamefont {T.}~\bibnamefont
  {Stammbach}}\ and\ \bibinfo {author} {\bibfnamefont {R.~L.}\ \bibnamefont
  {Walter}},\ }\href {\doibase 10.1016/0375-9474(72)90166-2} {\bibfield
  {journal} {\bibinfo  {journal} {Nuclear Physics A}\ }\textbf {\bibinfo
  {volume} {180}},\ \bibinfo {pages} {225 } (\bibinfo {year}
  {1972})}\BibitemShut {NoStop}%
\bibitem [{\citenamefont {Cs\'ot\'o}\ and\ \citenamefont {Hale}(1997)}]{Cso97}%
  \BibitemOpen
  \bibfield  {author} {\bibinfo {author} {\bibfnamefont {A.}~\bibnamefont
  {Cs\'ot\'o}}\ and\ \bibinfo {author} {\bibfnamefont {G.~M.}\ \bibnamefont
  {Hale}},\ }\href {\doibase 10.1103/PhysRevC.55.536} {\bibfield  {journal}
  {\bibinfo  {journal} {Phys. Rev. C}\ }\textbf {\bibinfo {volume} {55}},\
  \bibinfo {pages} {536} (\bibinfo {year} {1997})}\BibitemShut {NoStop}%
\bibitem [{\citenamefont {Wiringa}\ \emph {et~al.}(1995)\citenamefont
  {Wiringa}, \citenamefont {Stoks},\ and\ \citenamefont {Schiavilla}}]{Wir95}%
  \BibitemOpen
  \bibfield  {author} {\bibinfo {author} {\bibfnamefont {R.~B.}\ \bibnamefont
  {Wiringa}}, \bibinfo {author} {\bibfnamefont {V.~G.~J.}\ \bibnamefont
  {Stoks}}, \ and\ \bibinfo {author} {\bibfnamefont {R.}~\bibnamefont
  {Schiavilla}},\ }\href {\doibase 10.1103/PhysRevC.51.38} {\bibfield
  {journal} {\bibinfo  {journal} {Phys. Rev. C}\ }\textbf {\bibinfo {volume}
  {51}},\ \bibinfo {pages} {38} (\bibinfo {year} {1995})}\BibitemShut {NoStop}%
\bibitem [{\citenamefont {Bronson}(1964)}]{Bro64}%
  \BibitemOpen
  \bibfield  {author} {\bibinfo {author} {\bibfnamefont {J.~D.}\ \bibnamefont
  {Bronson}, \bibfnamefont {Jr.}},\ }\emph {\bibinfo {title} {Three-Alpha Decay
  of ${}^{12}{\rm C}$}},\ \href@noop {} {Ph.D. thesis},\ \bibinfo  {school}
  {Rice University} (\bibinfo {year} {1964})\BibitemShut {NoStop}%
\bibitem [{\citenamefont {Devons}\ and\ \citenamefont
  {Goldfarb}(1957)}]{Dev57}%
  \BibitemOpen
  \bibfield  {author} {\bibinfo {author} {\bibfnamefont {S.}~\bibnamefont
  {Devons}}\ and\ \bibinfo {author} {\bibfnamefont {L.~J.~B.}\ \bibnamefont
  {Goldfarb}},\ }in\ \href@noop {} {\emph {\bibinfo {booktitle} {Nuclear
  Reactions III, Encyclopedia of Physics, Volume 42}}},\ \bibinfo {editor}
  {edited by\ \bibinfo {editor} {\bibfnamefont {S.}~\bibnamefont {Fl{\"
  u}gge}}}\ (\bibinfo  {publisher} {Springer-Verlag},\ \bibinfo {address}
  {Berlin},\ \bibinfo {year} {1957})\ pp.\ \bibinfo {pages}
  {362--554}\BibitemShut {NoStop}%
\bibitem [{\citenamefont {Rose}(1958)}]{Ros58}%
  \BibitemOpen
  \bibfield  {author} {\bibinfo {author} {\bibfnamefont {M.~E.}\ \bibnamefont
  {Rose}},\ }\href@noop {} {\bibfield  {journal} {\bibinfo  {journal} {Journal
  of Mathematics and Physics}\ }\textbf {\bibinfo {volume} {37}},\ \bibinfo
  {pages} {215 } (\bibinfo {year} {1958})}\BibitemShut {NoStop}%
\bibitem [{\citenamefont {Bame}\ and\ \citenamefont {Leland}(1957)}]{Bam57}%
  \BibitemOpen
  \bibfield  {author} {\bibinfo {author} {\bibfnamefont {S.~J.}\ \bibnamefont
  {Bame}}\ and\ \bibinfo {author} {\bibfnamefont {W.~T.}\ \bibnamefont
  {Leland}},\ }\href {\doibase 10.1103/PhysRev.106.1257} {\bibfield  {journal}
  {\bibinfo  {journal} {Phys. Rev.}\ }\textbf {\bibinfo {volume} {106}},\
  \bibinfo {pages} {1257} (\bibinfo {year} {1957})}\BibitemShut {NoStop}%
\bibitem [{\citenamefont {Jarmie}\ and\ \citenamefont {Allen}(1958)}]{Jar58}%
  \BibitemOpen
  \bibfield  {author} {\bibinfo {author} {\bibfnamefont {N.}~\bibnamefont
  {Jarmie}}\ and\ \bibinfo {author} {\bibfnamefont {R.~C.}\ \bibnamefont
  {Allen}},\ }\href {\doibase 10.1103/PhysRev.111.1121} {\bibfield  {journal}
  {\bibinfo  {journal} {Phys. Rev.}\ }\textbf {\bibinfo {volume} {111}},\
  \bibinfo {pages} {1121} (\bibinfo {year} {1958})}\BibitemShut {NoStop}%
\bibitem [{\citenamefont {Brune}\ and\ \citenamefont {Sayre}(2013)}]{Bru13}%
  \BibitemOpen
  \bibfield  {author} {\bibinfo {author} {\bibfnamefont {C.~R.}\ \bibnamefont
  {Brune}}\ and\ \bibinfo {author} {\bibfnamefont {D.~B.}\ \bibnamefont
  {Sayre}},\ }\href {\doibase http://dx.doi.org/10.1016/j.nima.2012.09.023}
  {\bibfield  {journal} {\bibinfo  {journal} {Nuclear Instruments and Methods
  in Physics Research Section A: Accelerators, Spectrometers, Detectors and
  Associated Equipment}\ }\textbf {\bibinfo {volume} {698}},\ \bibinfo {pages}
  {49 } (\bibinfo {year} {2013})}\BibitemShut {NoStop}%
\bibitem [{\citenamefont {Jarmie}\ and\ \citenamefont {Brown}(1985)}]{Jar85}%
  \BibitemOpen
  \bibfield  {author} {\bibinfo {author} {\bibfnamefont {N.}~\bibnamefont
  {Jarmie}}\ and\ \bibinfo {author} {\bibfnamefont {R.~E.}\ \bibnamefont
  {Brown}},\ }\href {\doibase 10.1016/0168-583X(85)90279-4} {\bibfield
  {journal} {\bibinfo  {journal} {Nuclear Instruments and Methods in Physics
  Research Section B: Beam Interactions with Materials and Atoms}\ }\textbf
  {\bibinfo {volume} {10/11}},\ \bibinfo {pages} {405 } (\bibinfo {year}
  {1985})}\BibitemShut {NoStop}%
\bibitem [{\citenamefont {Jarmie}\ \emph {et~al.}(1984)\citenamefont {Jarmie},
  \citenamefont {Brown},\ and\ \citenamefont {Hardekopf}}]{Jar84}%
  \BibitemOpen
  \bibfield  {author} {\bibinfo {author} {\bibfnamefont {N.}~\bibnamefont
  {Jarmie}}, \bibinfo {author} {\bibfnamefont {R.~E.}\ \bibnamefont {Brown}}, \
  and\ \bibinfo {author} {\bibfnamefont {R.~A.}\ \bibnamefont {Hardekopf}},\
  }\href {\doibase 10.1103/PhysRevC.29.2031} {\bibfield  {journal} {\bibinfo
  {journal} {Phys. Rev. C}\ }\textbf {\bibinfo {volume} {29}},\ \bibinfo
  {pages} {2031} (\bibinfo {year} {1984})}\BibitemShut {NoStop}%
\bibitem [{\citenamefont {Alt}\ and\ \citenamefont
  {Mukhamedzhanov}(1993)}]{Alt93}%
  \BibitemOpen
  \bibfield  {author} {\bibinfo {author} {\bibfnamefont {E.~O.}\ \bibnamefont
  {Alt}}\ and\ \bibinfo {author} {\bibfnamefont {A.~M.}\ \bibnamefont
  {Mukhamedzhanov}},\ }\href {\doibase 10.1103/PhysRevA.47.2004} {\bibfield
  {journal} {\bibinfo  {journal} {Phys. Rev. A}\ }\textbf {\bibinfo {volume}
  {47}},\ \bibinfo {pages} {2004} (\bibinfo {year} {1993})}\BibitemShut
  {NoStop}%
\bibitem [{\citenamefont {Mukhamedzhanov}\ and\ \citenamefont
  {Lieber}(1996)}]{Muk96}%
  \BibitemOpen
  \bibfield  {author} {\bibinfo {author} {\bibfnamefont {A.~M.}\ \bibnamefont
  {Mukhamedzhanov}}\ and\ \bibinfo {author} {\bibfnamefont {M.}~\bibnamefont
  {Lieber}},\ }\href {\doibase 10.1103/PhysRevA.54.3078} {\bibfield  {journal}
  {\bibinfo  {journal} {Phys. Rev. A}\ }\textbf {\bibinfo {volume} {54}},\
  \bibinfo {pages} {3078} (\bibinfo {year} {1996})}\BibitemShut {NoStop}%
\bibitem [{\citenamefont {Wong}\ \emph {et~al.}(1965)\citenamefont {Wong},
  \citenamefont {Anderson},\ and\ \citenamefont {McClure}}]{Won65}%
  \BibitemOpen
  \bibfield  {author} {\bibinfo {author} {\bibfnamefont {C.}~\bibnamefont
  {Wong}}, \bibinfo {author} {\bibfnamefont {J.~D.}\ \bibnamefont {Anderson}},
  \ and\ \bibinfo {author} {\bibfnamefont {J.~W.}\ \bibnamefont {McClure}},\
  }\href {\doibase 10.1016/0029-5582(65)90040-4} {\bibfield  {journal}
  {\bibinfo  {journal} {Nuclear Physics}\ }\textbf {\bibinfo {volume} {71}},\
  \bibinfo {pages} {106 } (\bibinfo {year} {1965})}\BibitemShut {NoStop}%
\bibitem [{Zyl()}]{Zyl14}%
  \BibitemOpen
  \href@noop {} {}\bibinfo {note} {{A.B. Zylstra (unpublished)}.}\BibitemShut
  {Stop}%
\bibitem [{\citenamefont {Spyrou}\ \emph {et~al.}(2012)\citenamefont {Spyrou},
  \citenamefont {Kohley}, \citenamefont {Baumann}, \citenamefont {Bazin},
  \citenamefont {Brown}, \citenamefont {Christian}, \citenamefont {DeYoung},
  \citenamefont {Finck}, \citenamefont {Frank}, \citenamefont {Lunderberg},
  \citenamefont {Mosby}, \citenamefont {Peters}, \citenamefont {Schiller},
  \citenamefont {Smith}, \citenamefont {Snyder}, \citenamefont {Strongman},
  \citenamefont {Thoennessen},\ and\ \citenamefont {Volya}}]{Spy12}%
  \BibitemOpen
  \bibfield  {author} {\bibinfo {author} {\bibfnamefont {A.}~\bibnamefont
  {Spyrou}}, \bibinfo {author} {\bibfnamefont {Z.}~\bibnamefont {Kohley}},
  \bibinfo {author} {\bibfnamefont {T.}~\bibnamefont {Baumann}}, \bibinfo
  {author} {\bibfnamefont {D.}~\bibnamefont {Bazin}}, \bibinfo {author}
  {\bibfnamefont {B.~A.}\ \bibnamefont {Brown}}, \bibinfo {author}
  {\bibfnamefont {G.}~\bibnamefont {Christian}}, \bibinfo {author}
  {\bibfnamefont {P.~A.}\ \bibnamefont {DeYoung}}, \bibinfo {author}
  {\bibfnamefont {J.~E.}\ \bibnamefont {Finck}}, \bibinfo {author}
  {\bibfnamefont {N.}~\bibnamefont {Frank}}, \bibinfo {author} {\bibfnamefont
  {E.}~\bibnamefont {Lunderberg}}, \bibinfo {author} {\bibfnamefont
  {S.}~\bibnamefont {Mosby}}, \bibinfo {author} {\bibfnamefont {W.~A.}\
  \bibnamefont {Peters}}, \bibinfo {author} {\bibfnamefont {A.}~\bibnamefont
  {Schiller}}, \bibinfo {author} {\bibfnamefont {J.~K.}\ \bibnamefont {Smith}},
  \bibinfo {author} {\bibfnamefont {J.}~\bibnamefont {Snyder}}, \bibinfo
  {author} {\bibfnamefont {M.~J.}\ \bibnamefont {Strongman}}, \bibinfo {author}
  {\bibfnamefont {M.}~\bibnamefont {Thoennessen}}, \ and\ \bibinfo {author}
  {\bibfnamefont {A.}~\bibnamefont {Volya}},\ }\href {\doibase
  10.1103/PhysRevLett.108.102501} {\bibfield  {journal} {\bibinfo  {journal}
  {Phys. Rev. Lett.}\ }\textbf {\bibinfo {volume} {108}},\ \bibinfo {pages}
  {102501} (\bibinfo {year} {2012})}\BibitemShut {NoStop}%
\bibitem [{\citenamefont {Barker}(2003)}]{Bar03}%
  \BibitemOpen
  \bibfield  {author} {\bibinfo {author} {\bibfnamefont {F.~C.}\ \bibnamefont
  {Barker}},\ }\href {\doibase 10.1103/PhysRevC.68.054602} {\bibfield
  {journal} {\bibinfo  {journal} {Phys. Rev. C}\ }\textbf {\bibinfo {volume}
  {68}},\ \bibinfo {pages} {054602} (\bibinfo {year} {2003})}\BibitemShut
  {NoStop}%
\bibitem [{\citenamefont {Bartlett}\ \emph {et~al.}(2008)\citenamefont
  {Bartlett}, \citenamefont {Tostevin},\ and\ \citenamefont
  {Thompson}}]{Bar08}%
  \BibitemOpen
  \bibfield  {author} {\bibinfo {author} {\bibfnamefont {A.~J.}\ \bibnamefont
  {Bartlett}}, \bibinfo {author} {\bibfnamefont {J.~A.}\ \bibnamefont
  {Tostevin}}, \ and\ \bibinfo {author} {\bibfnamefont {I.~J.}\ \bibnamefont
  {Thompson}},\ }\href {\doibase 10.1103/PhysRevC.78.054603} {\bibfield
  {journal} {\bibinfo  {journal} {Phys. Rev. C}\ }\textbf {\bibinfo {volume}
  {78}},\ \bibinfo {pages} {054603} (\bibinfo {year} {2008})}\BibitemShut
  {NoStop}%
\end{thebibliography}%

\end{document}